\RequirePackage{rotating}
\documentclass[fleqn,useams,usenatbib]{mnras}
\usepackage{newtxtext,newtxmath}
\usepackage{multirow}
\usepackage{url}
\usepackage{hyperref}
\usepackage[T1]{fontenc}
\usepackage{ae,aecompl}
\usepackage{graphicx}
\usepackage{threeparttable}
\usepackage{booktabs}

\usepackage{wasysym} 
\usepackage{scalerel} 
\usepackage{verbatim} 
\usepackage{xcolor}
\usepackage{listings}

\newcommand{\msun}{h^{-1}\,{\rm M}_\odot}

\newcommand{\lm}{\log\,(m/{\rm M}_\odot)}

\definecolor{grey}{rgb}{0.4,0.5,0.6}
\definecolor{brown}{rgb}{0.65,0.16,0.16}
\definecolor{darkgreen}{rgb}{0.0,0.45,0.0}
\definecolor{darkorange}{rgb}{0.9,0.2,0.0}

\defcitealias{DiazGimenez+20}{Paper~I}
\defcitealias{DiazGimenez+21}{Paper~II}
\defcitealias{Taverna+22}{Paper~III}
\begin{document}

\title[Galaxy evolution in CGs with different assembly history]
{Compact groups from semi-analytical models of galaxy formation -- IV:\\
   effect of group assembly on the evolution of their galaxies}
\author[A. Zandivarez et al.]
{A. Zandivarez$^{1,2}$\thanks{ariel.zandivarez@unc.edu.ar}, 
E. D\'iaz-Gim\'enez$^{1,2}$,
A. Taverna$^{1,2}$,
G. A. Mamon$^{3}$
\\
\\
$^{1}$ Universidad Nacional de C\'ordoba (UNC). Observatorio Astron\'omico de C\'ordoba (OAC). C\'ordoba, Argentina\\
$^{2}$ CONICET. Instituto de Astronom\'ia Te\'orica y Experimental (IATE). C\'ordoba, Argentina \\
$^{3}$ Institut d'Astrophysique de Paris (UMR 7095: CNRS \& Sorbonne Universit\'e), Paris, France}

\date{Accepted XXX. Received YYY; in original form ZZZ}
\pubyear{2023}
\label{firstpage}
\pagerange{\pageref{firstpage}--\pageref{lastpage}}
\maketitle

\begin{abstract}
Using over 3000 compact groups (CGs) of galaxies extracted from mock catalogues built from semi-analytical models of galaxy formation (SAMs), we study whether the CG assembly channel affects the $z$=0 properties of galaxies and their evolution.  
The evolution of CG galaxy properties with time is a clear function of their stellar masses. For instance, high-stellar-mass CG galaxies have lived their last 8 Gyr with little cold gas content while maintaining their reservoir of hot gas, while low-mass CG galaxies still preserve some of their cold gas content at the present but they have completely drained their hot gas reservoir.
Beyond that, we find that the evolution of CG galaxies is also a function of the assembly history of the CGs: with more extreme losses of gas content, faster mass gain rates for black holes and more marked suppression of star formation as a function of cosmic time as we go from recent to early CG assembly. Thus, CGs constitute another laboratory for galaxy assembly bias, as the later assembling groups have later star formation. Our results show that classifying CGs according to their assembly channel is a way of distinguishing different paths by which galaxies transform their properties throughout their history. 
\end{abstract}


\section{Introduction}
\message{*** Intro ***}
A recurring topic during the last decades has been the study of the evolution of the properties of galaxies to try to understand the mechanisms by which these systems lose their ability to form stars at the present time. This suppression of star formation in galaxies, usually known as "quenching",  is one of the main topics in the current literature on extragalactic astronomy and is highly dependent on the galaxy's physical properties and the places where they inhabit. 

Analysing the $K$-band galaxy luminosity function, \cite{cowie+96} showed that galaxies undergoing rapid star formation show a declining luminosity with decreasing redshift. They suggested that this result is evidence that an important fraction of all galaxy formation occurred at early times and also that more massive galaxies formed at higher redshifts, naming this scenario \emph{downsizing}. This interpretation seems counter-intuitive to what the hierarchical structure formation scenario would indicate.
\cite{brinchmann-ellis+00} added more observational evidence that the most massive systems have already carried out most of their star formation at early times, while the less massive systems still show star formation events at present. They also suggested that the transformation of spiral galaxies into elliptical galaxies is unlikely to be a major process during the last 8 Gyr. 
\cite{thomas+10} analysed more than 3000 early-type galaxies in the Sloan Digital Sky Survey (hereafter, SDSS) Data Release 4 and concluded that the relevance of environmental processes increases with decreasing galaxy mass. Early-type galaxies are driven only by self-regulated processes mainly related to their mass, but in recent times other environmental processes such as tidal interactions and minor mergers have become important, producing a rejuvenation of the star formation in intermediate and low-mass early-type galaxies inhabiting low-density environments. 

At the same time, a similar result was obtained by \cite{peng+10}, who introduced the terms of mass and environmental quenching to explain the different processes that contribute to shaping the galaxy stellar mass function. 
Using galaxy groups in the SDSS Data Release 7 \citep{sdssdr7}, \cite*{wetzel+12} reported the bimodality of the specific star formation rate with a break at $10^{-11}\,\rm yr^{-1}$. They used this value to define as quenched those galaxies that lie below that cut and found that satellites are more prone to be quenched than central galaxies and the likelihood of being quenched increases with halo-centric radius and mass. 
On the other hand, \cite*{peng+15} used the stellar metallicity versus stellar mass to argue that local passive gas-poor galaxies of intermediate and low masses are most likely quenched by strangulation, which is described as a sudden interruption of cold gas availability. Observational evidence presented by \cite{Chen+20} showed that their new empirical model for the growth of black holes in central star-forming galaxies leads to a scenario where galaxy quenching is most likely due to AGN feedback.

As a crucial tool to understand the physics behind these findings, semi-analytical models (SAMs) of galaxy formation and evolution have provided valuable insights into galaxy evolution. There are various processes modelled in SAMs that can explain how galaxies are quenched in the Universe, including feedback processes due to the supernova (SN) explosions and active galactic nuclei (AGN) feedback from supermassive black holes in the centre of galaxies. Additionally, environmental processes caused by galaxy interactions with each other or with the gaseous medium within haloes are also relevant. One of the most hotly debated topics in the literature is determining which of these processes is most relevant in understanding how star formation is suppressed in the lifespan of galaxies. 

Applying various recipes from previous SAMs, \cite{delucia+12} demonstrated that AGN feedback effectively replicates the observed stellar population in massive galaxies. However, it falls short of reproducing the observed chemical abundances. Interestingly, even without incorporating AGN feedback, there was a significant reduction in the observed discrepancies.
After testing various feedback recipes, \cite*{hirschmann+16} found that the stellar and gas metallicity of galaxies can be reproduced using strong ejective feedback with a mass-loading that varies with redshift. 
In a study by \cite{bluck+16}, the Illustris hydrodynamical simulation \citep{illustris} and the Munich SAM (known as L-Galaxies) were compared. They found that central galaxies are quenched through AGN feedback. But the models don't match observations, since L-Galaxies quenching is too efficient while in Illustris is not efficient enough. 

\citeauthor{bluck+16} also observed that the environment affects the quenched fraction of lower-mass satellites while higher-mass satellites and centrals quench similarly. \cite{henriques+17} applied a SAM to the Millennium Simulation to show that AGN feedback is the main responsible for the quenching of massive galaxies, supporting previous work \citep{Silk&Rees98,croton+06}. \cite{henriques+17} also observed that both AGN feedback and environmental processes are enhanced in high-density environments. Environmental processes such as ram-pressure \citep{gunngott+72} and tidal stripping of gas \citep{merritt+83} are relevant only for the quenching of satellite galaxies (e.g, \citealt{jaffe+15,roberts+19,moretti+22}). 

\cite{cora+18} implemented on the SAG semi-analytical model a gradual removal of hot gas and, eventually, also cold gas of the disc (due to ram-pressure and tidal stripping in satellite galaxies) obtaining fractions of quenched galaxies consistent with observations. Then, \cite{cora+19} implemented a gradual starvation scenario into SAG, showing that a pronounced reduction of the cooling rate is sufficient to start the galaxy quenching. At about the same time, \cite*{delucia+19} showed that the SN feedback with strong stellar-driven outflows reproduces the variations of the fraction of quenched galaxies as a function of mass. 
 More recently, \cite{hough+23} implemented the SAG model on galaxy clusters and showed that roughly 85\% of the quenched population at present inside clusters are ancient infallers with almost zero hot and cold gas content at $z=0$. They argued that ram-pressure stripping is the main mechanism to remove hot gas after the galaxies have reached their first or second close passage through the centre of the cluster. 

The preceding paragraphs show just a few examples of the extensive research that has been conducted in recent years on the evolution of galaxy star formation in various environments. Such research has been approached from both observational and numerical perspectives, and the debate on the topic remains ongoing.
There are several environments in the Universe that are interesting for analysing the rapidity of galaxy evolution. One of these environments is compact groups of galaxies (CGs). These extremely dense, presumably isolated, systems of a few similar-luminosity galaxies are thought to constitute the environment where galaxy interactions and mergers are the most frequent \citep{Mamon92}. They became popular after the first systematic identification on observational data by \cite{Hickson82}, and have generated countless studies and controversies. One of the main concerns has been how these small systems have formed and for how long they have existed. Some authors have questioned their mere existence, arguing that such dense systems should see their galaxies rapidly merge into a single large galaxy \citep*{Carnevali+81,Barnes85,Navarro+87,Mamon87}. Others have argued that some of these dense groups could survive this merging instability (\citealt*{Diaferio+94} for groups dominated by a massive binary in a circular orbit and \citealt*{Athanassoula+97} for a uniform density group). 

\begin{table}
\setlength{\tabcolsep}{2pt}
\small{
\begin{center} 
\caption{Munich semi-analytical models analysed in this work \label{tab:sams}}
\begin{tabular}{clccccccccc}
\hline
\hline
\multicolumn{1}{c}{SAM} &\multicolumn{5}{c}{Simulation} & &  \multicolumn{3}{c}{Lightcone} \\ 
\cline{2-6} \cline{8-10}
acronym & \multicolumn{1}{c}{name} & $\Omega_{\rm m}$ & $h$ & $\sigma_8$ & box size && \# & CGs & CG4s\\
\multicolumn{1}{c}{(1)\ \ } & \multicolumn{1}{c}{(2)} &\multicolumn{1}{c}{(3)} & (4) & (5) & (6) && (7) & (8) & (9) \\
\hline
G11 & \ WMAP1 & 0.25 & 0.73 & 0.90 &  500  & & 3\,149\,024 & 3175 & 1571\\
G13 & \ WMAP7 & 0.27 & 0.70 & 0.81 &  500 & & 2\,982\,462 & 1682 & 1010\\
H20 & \ Planck & 0.31 & 0.67 & 0.83 & 480 & & 3\,491\,251 & 1895 & 1235\\
A21 & \ Planck & 0.31 & 0.67 & 0.83 & 480 & & 2\,892\,023 & \ \,758 & \ \,546\\
\hline
\end{tabular}  
\end{center} 
\parbox{\hsize}{\noindent Notes:
The columns are:
(1): SAM: G11
\citep{Guo+11}, G13 \citep{Guo+13}, H20 \citep{Henriques+20} and A21 \citep{Ayromlou21}; 
(2): cosmology of parent simulation;
(3): density parameter of parent simulation;
(4): dimensionless $z$=0 Hubble constant of parent simulation;
(5): standard deviation of the (linearly extrapolated to $z$=0) power spectrum on the scale of $8\,h^{-1}\,\rm Mpc$;
(6): periodic box size of parent simulation [$\, h^{-1} \, \rm Mpc$];
(7): number of galaxies with stellar masses greater than $7\times 10^8 \, \msun$ in the lightcone ($r\le 17.77$). 
(8): number of compact groups identified in the mock lightcones;
(9): number of compact groups with exactly four galaxy members.\\
More accurate values for the simulations can be found at
 \url{http://gavo.mpa-garching.mpg.de/Millennium/Help/simulation}.}
\parbox{\hsize}{\noindent  $^{*}$ H20 and A21 have smaller box sizes since they were run on a re-scaled version of the original Millennium simulation.}}
\end{table}

To unravel the assembly history of CGs, it is essential to use numerical simulations in a cosmological context, such as those performed using mock galaxy catalogues constructed using semi-analytical models of galaxy formation (SAMs) on top of $N$-body cosmological simulations (see for instance, \citealt*{McConnachie+08}; \citealt{DiazGimenez&Mamon10,DiazGimenez+15,Taverna+16}; \citealt*{DiazGimenez+18}; \citealt{Farhang+17,wiens+19}), or using hydrodynamical simulations \citep{Hartsuiker&Ploeckinger20}.
In recent years, we have embarked on a comprehensive study of CGs identified in several mock catalogues constructed from different SAMs. In \cite{DiazGimenez+20} (hereafter \citetalias{DiazGimenez+20}), we discovered a strong dependence of the frequency and nature of CGs on the cosmological parameters, as well as, to a lesser extent, on the SAM recipes. In \cite{Taverna+22} (hereafter \citetalias{Taverna+22}), we proposed a method to identify isolated dense groups in 3D real space that reproduce the compactness of the original visual Hickson CG sample. But it turns out that the Hickson-like automatic algorithm performs poorly in recovering these systems in a flux-limited catalogue, mainly because they are not truly isolated in 3D (isolation in redshift space rarely leads to isolation in real space). 

\cite*{DiazGimenez+21} (hereafter  \citetalias{DiazGimenez+21}) analysed the assembly history of Hickson-like CGs of four galaxies, and we have identified four different assembly channels for the formation of CGs. We found that only a few per cent of CGs assembled at early epochs, 10 to 20 per cent have formed by gradual contraction of their orbits during the last 7 Gyr, while most of the CGs assembled recently with the first or second arrival of the fourth galaxy. By analysing the CG observational properties at the present time ($z=0$), we also found that CGs characterised by an early assembly are more evolved, as they appear to be smaller in size as well as with more dominant first-ranked galaxies (luminous or most massive).

The extreme environment observed in CGs could provide different insights into how galaxies suppress their star formation over time. Therefore, in the present article, we extend \citetalias{DiazGimenez+21} by analysing the evolution of the galaxy properties in CGs when the different assembling channels are considered. We intend to disentangle how the evolution of galaxy parameters in CGs is affected by stellar mass, cosmological parameters, SAM recipes, and, for the first time, CG assembly histories.  
To perform this study, we used four mock CG samples built from the Millennium Simulations \citep{Springel+05}, with L-Galaxies SAMs previously used in \citetalias{DiazGimenez+20} and \citetalias{Taverna+22}. The SAMs were based on parent cosmological dark matter simulations using a wide range of cosmological parameters. Using the galaxy merger trees from the SAMs, we used the procedure defined in \citetalias{DiazGimenez+21} to classify CGs into the four different assembly channels. 

The layout of this work is as follows. In Sect.~\ref{sec:sample}, we present the SAMs and the lightcone catalogues built from these SAMs, and the mock CG samples used in this work. In Sect.~\ref{sec:histories}, we describe the classification of CGs according to their assembly channel. In Sect.~\ref{sec:stats}, we compare the $z$=0 CG galaxy properties as a function of SAM and CG assembly history and then analyze the evolution of CG galaxy properties again as a function of SAM and CG assembly history. We conclude and discuss our results in Sect.~\ref{sec:discus}.

\begin{figure*}
\begin{center}
\includegraphics[width=0.95\hsize]{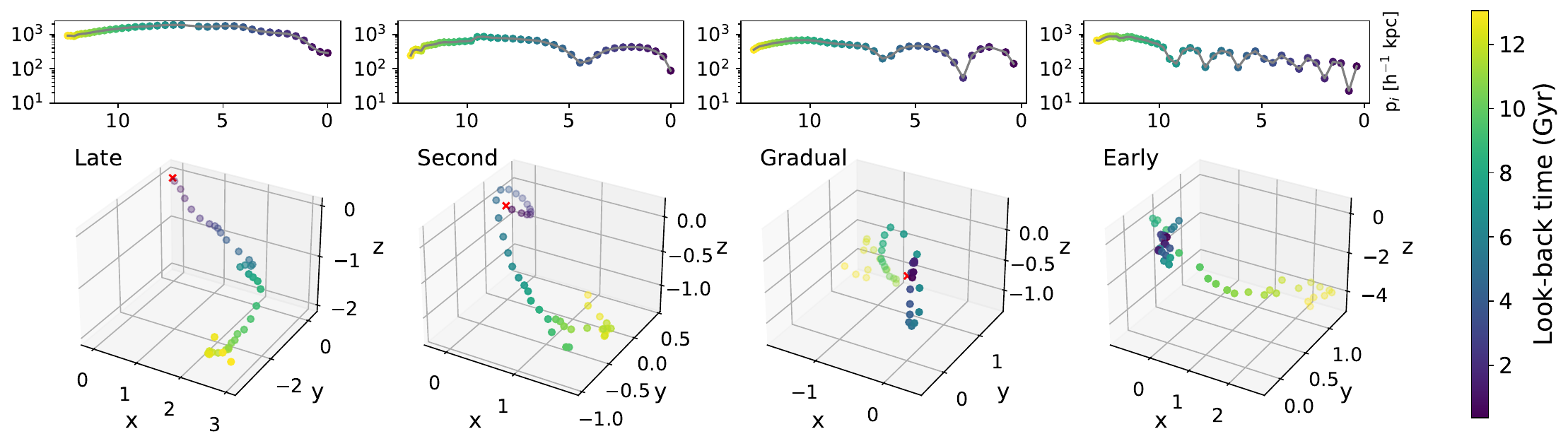}
\caption{
Examples of four CG4s with different assembly channels for the A21 SAM.
From \emph{left} to \emph{right}, we show CG4s classified as {\tt Late}, {\tt Second}, {\tt Gradual} and {\tt Early} assembly, respectively.
{\bf Upper panels}: 3D physical distance of the key galaxy member to the CG centre of mass as a function of look-back time. 
{\bf Lower plots}: orbits around the CG4 centre of mass described by the key galaxies in 3D space. Colours indicate the corresponding time (look-back time) for each snapshot of the simulation (darker colours towards the present time). The \emph{red cross} indicates the 3D position of the centre of mass, at all epochs. 
\label{fig:channels}}
\end{center}
\end{figure*}

\section{The Samples}
\label{sec:sample}
In this section, we briefly describe the mock lightcones built for this work and the procedure to identify CGs.

\subsection{Mock galaxy lightcones}
We used four L-Galaxies SAMs run on the Millennium simulation I \citep{Springel+05}: that of \cite{Guo+11} (hereafter G11) based on cosmological simulations in the WMAP1 cosmology \citep{Spergel+03}, \cite{Guo+13} (hereafter G13) in the WMAP7 cosmology \citep{Komatsu+11}, and the SAMs of \cite{Henriques+20} and \cite{Ayromlou21} (hereafter H20 and A21, respectively)  run on the Millennium simulation re-scaled to the Planck cosmology \citep{Planck+16}. Table~\ref{tab:sams} lists the different SAMs with their cosmological parameters and parent simulations. We only considered galaxies with stellar masses greater than $\sim 10^9 \, {\rm M}_\odot$ \citep{Guo+11,knebe15,irodotou19}. 

For each SAM listed in Table~\ref{tab:sams}, we constructed all-sky mock galaxy lightcones following a similar procedure as \cite{jpas}, previously described in \citetalias{DiazGimenez+20}. 
Briefly, we create lightcones using synthetic galaxies taken from different redshift slices in the simulation outputs at various redshifts. The width of the slices is determined by the separation between snapshots in the simulation. This method enables us to track the changes in structures and galaxy properties over time, while also accounting for any missing or duplicated galaxies due to movements between snapshots. The SAMs provide us with rest-frame galaxy absolute magnitudes, which we interpolate between different snapshots. To determine the observer-frame apparent magnitudes, we follow the k-decorrection procedure described in \cite{DiazGimenez+18}.
These lightcones are limited to an apparent observer-frame Sloan Digital Sky Survey (SDSS) AB magnitude of $r \leq 17.77$. The number of galaxies in the lightcone of each SAM is quoted in Table~\ref{tab:sams}.

\begin{table}
\tabcolsep 2.5pt
\centering
\caption{Number of 4-galaxy CGs classified by assembly channel}
\begin{tabular}{lrrrrrr}
\hline
SAM & CG4 & Late & Second & Gradual & Early & Fake \\
\hline
All & 4362 & 1633 & 1082 & 592 & 217 & 838 \\
\hline
G11 & 1571 & 541 & 417 & 244 & 119 & 250 \\
G13 & 1010 & 402 & 228 & 149 &  29 & 202 \\
H20 & 1235 & 464 & 315 & 148 &  57 & 251 \\
A21 &  546 & 226 & 122 &  51 &  12 & 135 \\
\hline
\hline
\end{tabular}
\parbox{\hsize}{We also quote, in the last column, the number of CG4s considered as Fake associations in 3D real space.}
\label{tab:asch}
\end{table}

\subsection{Hickson-like compact group identification}

We identify CGs using the CG finding algorithm of \cite{DiazGimenez+18} (as in \citetalias{DiazGimenez+20}). 
The finder looks for Hickson-like groups that simultaneously satisfy constraints on membership, flux limit of the brightest galaxy, compactness of the group, relative isolation and velocity concordance. 
Particularly, the following criteria must be satisfied: 
\begin{description}
    \item[Population:] The number of bright members (in a 3-magnitude range from the brightest) is 4 or more galaxies
    \item[Flux limit of the BGG:] The brightest galaxy of the system has to be at least 3 magnitudes brighter than the catalogue magnitude limit.
    \item[Compactness:] the surface brightness of the group in the $r$-band is less than $26.33 \, \rm mag \, arcsec^{-2}$.
    \item[Isolation:] There are no other bright galaxies (in a 3-mag range from the brightest, nor brighter) within an angular radius of three times the radius of the minimum circle that encloses the galaxy members.
    \item[Velocity concordance:] all the galaxy members are within $1\,000 \rm \, km\,s^{-1}$ from the median velocity of the group. 
\end{description}

When working with semi-analytical galaxies, the algorithm also takes into account the possible blending of observational galaxies. To determine the size of the galaxies in the lightcones, we assigned a half-light radius based on their stellar masses, following the recipes in \cite{Lange+15}. We then identified two galaxies as blended if their projected radii overlapped on the sky plane. This blending affects the membership of the groups.
Following \citetalias{DiazGimenez+21}, to simplify the analysis, we adopt for this work only those CGs with four galaxies in a 3-magnitude range (hereafter, CG4s). In Table~\ref{tab:sams}, the number of CG4s for each SAM is quoted. They represent between $50$ to $70$ per cent of each CG sample.

It has been previously noted \citep{McConnachie+08,DiazGimenez&Mamon10} that the Hickson-like algorithms identify a non-negligible fraction of systems that are not physically dense aggregates, but chance alignments along the line of sight. 
Using real-space information from the simulations, we then exclude systems with maximum 3D inter-galaxy separations larger than $1\,h^{-1} \rm Mpc$ at 
 present time, and those without close pairs in 3D space (i.e., without galaxies separated by less than $200\,h^{-1}\,\rm kpc$). These particular types of CG4s (named ``Fakes'' in \citetalias{DiazGimenez+21}) will not be considered in this work.

\section{CG classification according to their assembly history}
\label{sec:histories}

\message{*** Histories ***}
CG4s are classified into different assembly channels by following the procedure devised in \citetalias{DiazGimenez+21}.
Briefly, we track the evolutionary path of each $z$=0 galaxy member of the CG4 sample using their merger tree histories\footnote{Merger histories are retrieved from \url{http://gavo.mpa-garching.mpg.de/Millennium/}}.
We only follow the position of the main progenitor of each $z$=0 galaxy up to the redshift where the galaxy is formed. 
For each snapshot $i$ at look-back time $t_i$ (where $i$ runs from 0 to the total number of snapshots), we calculate the physical 3D distance, $p_j(t_i)$ between each galaxy, $j$, and the stellar mass centre of the group\footnote{ 
$p_{j}(t_i) = d_j(t_i)/(1\!+\!z_i)$, 
where $d_j(t_i)$ is the modulus of the 3D comoving distance to the group centre}. Using the $p_j(t_i)$ vs $t_i$ profiles we measure three quantities\footnote{see \citetalias{DiazGimenez+21} for computational details}: 
\begin{enumerate}
\item $t_{\rm 1p}$: time of the first (earliest) pericentre for galaxy $j$;
\item $n_{\rm p}$: total number of deep pericentres\footnote{ 
A pericenter is when the slope of the profile changes from negative to positive. To be considered a deep pericenter at $t_i$, the height of the profile has to be less than 0.8 of the height at $t_{i-2}$ and $t_{i+2}$. 
} for galaxy $j$;
\item $p_j(t_0)/p_j(t_1)$: ratio between the $p_j$ in $t_0$ (snapshot at $z=0$) and the immediately subsequent output, $t_1$.
\end{enumerate}
In \citetalias{DiazGimenez+21}, we defined four different types of CG assembly channels by analysing the behaviour of the galaxy member of each CG4 with the latest arrival to the group (named as ``key'' galaxy). The key galaxy is the member with the fewest $n_{\rm p}$. 
To decide between more than one galaxy with the same lowest values of $n_{\rm p}$, we have chosen the galaxy with the most recent $t_{\rm 1p}$ provided that the $t_{\rm 1p}$ of those galaxies are separated for more than 1 Gyr. If this requirement is not met by either galaxy, the key galaxy is the one with the highest value of $p_j(t_{1p})$ at its first pericenter. Using the selected key galaxy, in \citetalias{DiazGimenez+21}, we have devised 
a set of criteria to associate a CG4 with a corresponding assembly channel. 
Hence, a CG4 could have: 
\begin{itemize}
\itemindent=0pt
\item  {\tt Late Assembly}, hereafter {\tt Late}: the fourth galaxy has just arrived in the CG for the first time, when 
$t_{\rm 1p}<7.5$ Gyr AND \{$n_{\rm p}=0$ OR [$n_{\rm p}=1$ AND $p_{\rm key}(t_0)/p_{\rm key}(t_1) >1$]\}.
\item {\tt Late Second Pericentre}, hereafter {\tt Late}: the fourth galaxy has arrived in the CG on its second passage
(\texttt{Second}), when $t_{\rm 1p}<7.5$ Gyr AND 
\{[$n_{\rm p}=1$ AND $p_{\rm key}(t_0)/p_{\rm key}(t_1) \leq 1$] OR [$n_{\rm p}=2$ AND 
$p_{\rm key}(t_0)/p_{\rm key}(t_1) > 1$]\}.
\item {\tt Gradual Contraction}, hereafter {\tt Gradual}: the four galaxies might or might not have been together since an early epoch. Still, they have already completed two or more orbits together, becoming gradually closer with each orbit (\texttt{Gradual}). This behaviour is described when $t_{\rm 1p}<7.5$ Gyr AND \{[$n_{\rm p}=2$ AND $p_{\rm key}(t_0)/p_{\rm key}(t_1) \leq 1$] OR $n_{\rm p}>2$\}.
\item {\tt Early Assembly}, hereafter {\tt Early}: the four galaxies have been together since an early epoch (hereafter \texttt{Early}) when $ t_{\rm 1p} \ge 7.5$ Gyr.
\end{itemize}

The application of this 
classification in \citetalias{DiazGimenez+21} led to a 96 per cent success at recovering a previous visual classification of a test sample. 
In Fig.~\ref{fig:channels}, we display examples of the orbits of the key galaxy of four CG4s with different assembly channels. Upper panels show the physical distance of the key galaxy to the mass centre as a function of look-back time, while lower plots show the positions of the key galaxy with respect to the group mass centre through time in 3D real space. 
In Table~\ref{tab:asch}, we quote the number of CG4s in each assembly channel for the four SAMs used in this work.
As previously observed in \citetalias{DiazGimenez+21}, $\sim 60\%$ of CG4s are characterised by the late assembly (Late+Second), while older systems comprise between 11 to 24\% of the CG4s (Gradual+Early). The remaining percentages correspond to Fake CGs. 
The lowest percentage of older and probably more evolved systems is obtained for the A21. 

In the SAMs used in this work, galaxies are classified into three different types according to the nature of their associated subhalo:  centrals, which reside in the central subhalo of a FOF halo;  satellites, with a subhalo which is not the central subhalo in a given FOF; and orphans, which are satellite galaxies whose associated subhalos have been tidally disrupted preventing their detection above a given threshold. Usually, the SAMs define the position and velocity of orphan galaxies by tracking the most bound particle of their disrupted subhalo (other SAMs use recipes for the positions of galaxies based on orbital decay by dynamical friction, which are uncertain for elongated orbits). Therefore, the possibility of a key galaxy (used to define assembly channels) being an orphan galaxy could raise some concerns. 
In \citetalias{DiazGimenez+20} we have shown that, on average, we could expect for a four-member CG that there is one central, one satellite and two orphans. In  \citetalias{DiazGimenez+21} we studied the percentage of key galaxies that are orphans and the distance from the CG centre at the moment when they become orphans. Our analysis showed that while some key galaxies are orphans, this does not have an impact on our results for how CG galaxies evolve according to the assembly channel of the CG.

\section{Results}
\label{sec:stats}

\message{*** Results ***}

\subsection{Galaxy properties at redshift zero}
\label{sec:z0}
In this section, we analyse the behaviour of fundamental CG galaxy properties at the present time. 
We have chosen properties that can be thought of as proxies for the efficiency of a galaxy to form stars, and that also provide information on the processes that regulate this efficiency.
Particularly, for each galaxy in CGs, we analyse its content of Cold Gas, Hot Gas, Black Hole mass and the star formation rate (SFR).

\begin{figure*}
    \begin{center}
    \includegraphics[width=0.9\hsize]{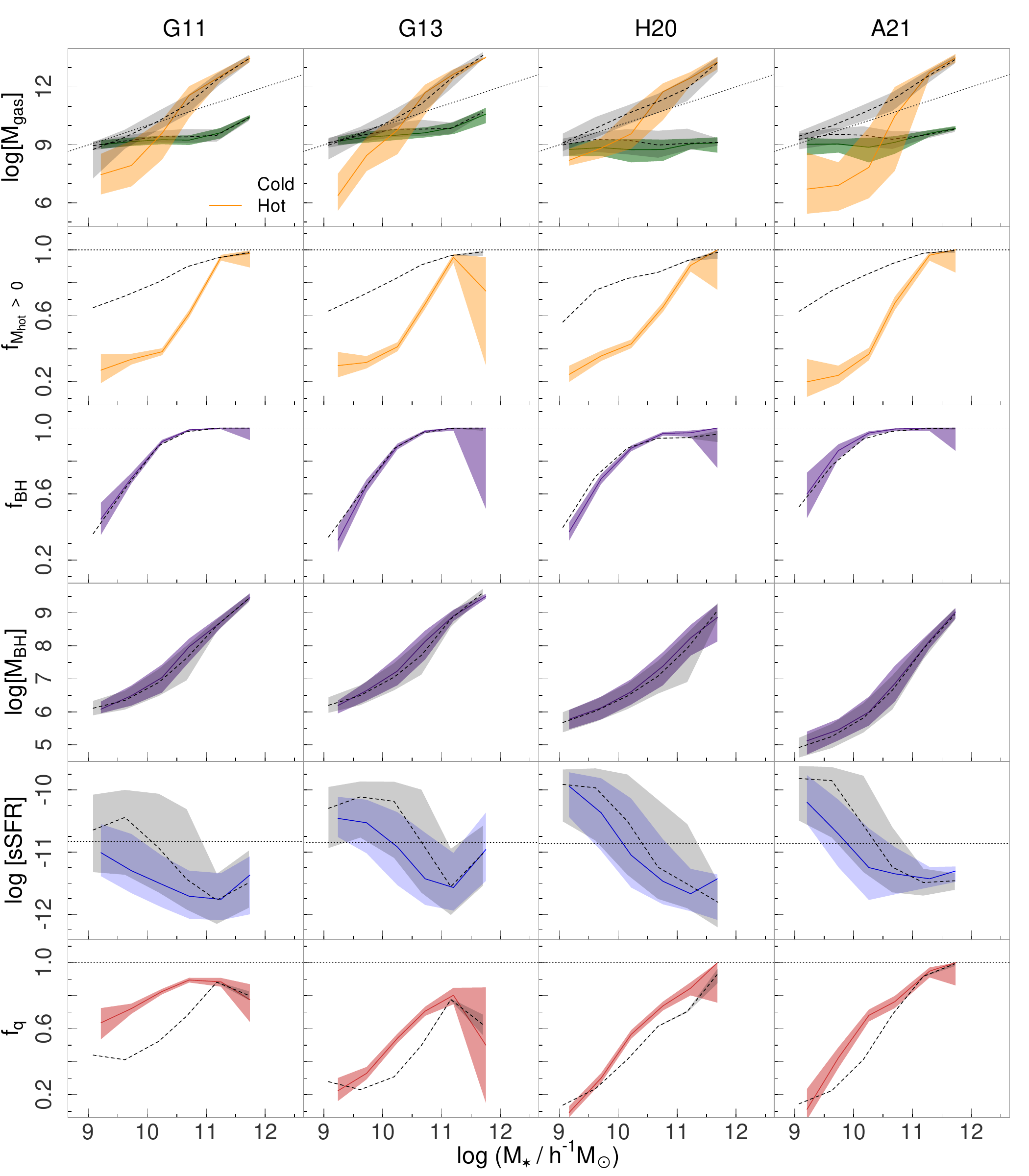}
    \caption{\label{fig:propsz0} Medians of the
    different galaxy properties as a function of their galaxy stellar mass both at present time ($z=0$). The distributions displayed in colours (solid lines) show galaxies in CGs while those in grey (dashed lines) are for the full sample of galaxies in the simulation box. 
    Galaxy properties are (from \emph{top} to \emph{bottom}) the cold and hot gas, the fraction of galaxies with hot gas, the black hole occupation fraction, the black hole masses, the specific SFR and the fraction of quenched galaxies. Each column shows the variation for a given SAM. In the first row, the \emph{dotted line} displays the identity relation for the two axes. The \emph{horizontal dotted lines} on the second, third, and bottom row panels represent the value of the fraction equal to one, while the \emph{horizontal dotted line} in the specific SFR indicates the limit below which galaxies can be quenched (see Section~\ref{sec:z0} for details). 
    The \emph{shaded regions} cover the interquartile range for the masses and sSFR, and the binomial 95\% confidence interval \protect{\citep{wilson27}} for the fractions.
    Mass units are in $\msun$, while sSFR units are in $\rm yr^{-1}$.
    }
    \end{center}
\end{figure*}

Several processes that govern galaxy formation and evolution are related to the total stellar mass of the galaxy. Figure~\ref{fig:propsz0} shows each galaxy property as a function of their total stellar mass, both at the present time. The properties of CG galaxies are shown with different colours (solid lines), while the grey lines (black dashed lines) display the trends for the full sample of galaxies in the $z=0$ snapshot of each SAM.

In general, hot gas masses for galaxies in CGs increase strongly with stellar mass and are very similar among the different SAMs. In this case, the hot gas content is $\sim 2$ dex below the lowest stellar masses and 1.5 dex above the highest stellar mass content, i.e, a total growth of $\sim$ 6 dex in the full range of stellar masses (whose range spans roughly half that size).
An interesting result is observed when comparing with the full sample of galaxies. For stellar masses lower than $10^{10.5} \,h^{-1} \,{\rm M}_\odot$, galaxies in CGs show less hot gas than that observed in the general population of galaxies. This behaviour is present regardless of the SAM, but the decrease is much more pronounced for G13 and A21 in the low stellar mass end. 
In the second row Fig.~\ref{fig:propsz0}, we show the fraction of galaxies with some content of hot gas mass at $z=0$. Regardless of the SAM, a large fraction ($\sim 70\%$) of low stellar mass galaxies (less than $10^{10} \, \msun$) in CGs have lost their hot gas content, while galaxies with the largest stellar masses (greater than $10^{11} \, \msun$) in this systems have a negligible fraction of galaxies without hot gas. This behaviour is very different to the observed for the full set of galaxies. The fraction trends are an increasing function with the stellar mass, but at the low mass end, only 40\% of the galaxies have lost their hot gas content. Therefore, it is very noticeable the depletion of hot gas for galaxies in CGs in the low stellar mass range compared with the general population of galaxies. This result combined with the obtained in the previous panel, shows that not only there are very few low mass galaxies in CGs with some hot gas content, but also their reservoir of hot gas is much smaller than expected for galaxies in general.

Also, in the first row of Fig.~\ref{fig:propsz0} we show the variation of the cold gas mass as a function of stellar mass. In this case, there is almost no difference when comparing galaxies in CGs with those in the full sample, except for a very small tendency for galaxies in CGs to have smaller cold gas masses in the low stellar mass range than in average for H20 and A21.
Considering only galaxies in CGs, in G11 and G13, for low stellar mass galaxies, the cold gas masses have similar values (green trends at the first row of Fig.~\ref{fig:propsz0}), while for high stellar mass galaxies, the cold gas masses are almost 1 dex smaller than the total stellar masses. A21 shows a slower growth rate of cold gas mass with stellar mass, and H20 has an almost constant behaviour regardless of their stellar content. 
A constant cold gas mass implies a cold gas to stars ratio that varies as $1/M_*$, which contrasts with the shallower slopes of $-0.75$ to $-0.5$ found in observations \citep{baldry08,Catinella+18}.

In the third row of Fig.~\ref{fig:propsz0}, we show the black hole occupation fraction ($f_{\rm BH}$) for each sample, i.e., the fraction of galaxies that actually host a black hole at z=0. We can see that above $10^{10} \ \msun$ the $f_{\rm BH}$ the fraction has a minimum of 0.9, increasing toward the highest stellar masses. Below that mass limit, the $f_{\rm BH}$ drops rapidly reaching values $\sim 0.3 -0.4$ for the lowest galaxy stellar masses. This behaviour is almost independent of the SAM analysed and displays almost no difference when comparing galaxies in CGs with those in the full sample of galaxies.

These results can be compared with the work of \cite{haidar+22}. They estimated the $f_{\rm BH}$ for a suite of hydrodynamical simulations (see \citealt{habouzit+21} for a summary of the simulations) and compared their results with observational data (see their fig.~1). In the regime of low stellar masses, the hydrodynamical simulations EAGLE \citep{schaye+15}, Illustris \citep{springel+10} and TNG \citep{springel+18}  show larger $f_{\rm BH}$ than found in the SAMs in this work, while SIMBA \citep{dave+19} and Horizon-AGN \citep{dubois+14} simulations display lower $f_{\rm BH}$ than the ones found here for the SAMs. At high stellar masses, there is excellent agreement between hydrodynamical simulations and the SAMs shown here. 
When comparing with observational data, the results of \cite{trump+15} and \cite{nguyen+18} are in better agreement with the $f_{\rm BH}$ obtained here for the SAMs.
The trends of the black hole masses (fourth row) are quite similar between the different SAMs, displaying an increasing function with 
the galaxy stellar mass spanning a range from $10^6$ to $10^{9.5}$ $\msun$ of black hole masses. Again, we see excellent agreement between galaxies in CGs and the complete sample of galaxies at $z=0$.

The fifth row of Fig.~\ref{fig:propsz0} shows the specific star formation rate (sSFR = SFR/$M_{*}$) as a function of the stellar mass of the galaxies. 
The general trend is an sSFR decreasing with the galaxy's stellar mass, i.e., galaxies are more likely to form stars at a higher rate when their stellar mass is low, while for high stellar masses, a lower star formation rate is expected. This general behaviour varies with the SAM. For G11, the decreasing behaviour with increasing mass is less pronounced, while in G13 the trend is steeper. For H20 and A21, the sSFR is in agreement with the general trend, i.e., a clear decreasing function with increasing stellar mass, spanning a wider range of sSFR values than in the G11 and G13. G11 and G13 display a rise of sSFR at the higher masses, which is less the case in H20 and A21. Such a rise in sSFR at high mass is not seen in observations (e.g., \citealt{Knobel+15}). Finally, for stellar masses lower than $10^{10.5} \,h^{-1} \,{\rm M}_\odot$ we see a shift toward lower sSFR values for galaxies in CGs compared to the full galaxy sample.

These estimates of the sSFR for galaxies in CGs can be used to study the fraction of galaxies with suppressed SFR. A galaxy is considered as with suppressed SFR (hereafter quenched galaxy) when its sSFR is below a certain limit. 
Over the years, several works have set this limit differently arguing that the correct value is such to split correctly the bimodal distribution of the sSFR. 
For instance, \cite{wetzel+12} and \cite{henriques+17} set $\rm \log_{10}[sSFR/(yr^{-1})]<-11$ at $z=0$ to define quenched galaxies, while \cite{brown+17} and \cite{cora+18} preferred a threshold of --10.7. 
Then, \cite{lacerna+22} preferred to set the limit to --10.5, while \cite{Henriques+20, Ayromlou21, Ayromlou+22} were prone to keep the value at --11 at $z=0$, as initially used. 
But what is also in common in these last three works, is that they vary this limit depending on the age of the Universe. 
Since the sSFR distribution changes in early times shifting the medians toward higher values, 
\cite{Henriques+20} defined a reasonable limit as a function of redshift could be 1 dex below the median of the sSFR main sequence, where the median can be roughly estimated by $2\, H_0 \,(1+z)^2$, with $H_0$ the Hubble constant at $z=0$. This recipe for the threshold, which at present time can be written as $0.2 \,h_0$, is similar to $0.3 \,h_0$ suggested by \cite{franx+08} ($\sim 0.15$ dex difference).
Therefore, in this work, we use \cite{Henriques+20} prescription to calculate the fraction of quenched galaxies as a function of time (redshift). Since we are dealing in this work with three different cosmologies, the sSFR limiting value should be obtained according to the value of the Hubble constant at $z=0$ (see Table~\ref{tab:sams}). 
Nevertheless, it is worth mentioning that a change of  10\% (0.04 dex) in the Hubble constant  produces variations in the limiting value of only 0.01 dex, hence there is almost no variation with cosmology. 
Instead, using this prescription, thresholds of roughly --10.9, --10.3 and --9.9 correspond to $z=0,1$ and 2, respectively (look-back times of approximately 0, 8 and 10 Gyr).

In the bottom row of Fig.~\ref{fig:propsz0}, we show the trends obtained for the fraction of quenched galaxies in CGs ($f_{\rm q}$). The $f_{\rm q}$ are estimated using the prescription of \cite{Henriques+20} for $z=0$ (these values are shown as dashed horizontal lines in the sSFR distributions of this figure). 
The difference observed in the sSFR has an implication on this fraction. There is a higher fraction of quenched galaxies in CGs than observed in the complete sample of galaxies in the $z=0$ snapshot (regardless of the SAM) for stellar masses lower than $10^{11} \,h^{-1} \,{\rm M}_\odot$.
From the analysis of the $f_{\rm q}$ for galaxies in CGs, we can see that for G13, H20 and A21 galaxies with masses larger than $10^{10} \, \msun$ have at least a 50\% chance of being quenched at $z=0$ (most galaxies with the highest stellar masses are certainly quenched). Galaxies with the lowest stellar masses are prone to remain active at the present time. Surprisingly, in contrast with observations \citep{Salim+07}, there is (marginal) evidence that the fraction of quenched galaxies decreases at the highest stellar masses for the G11 and G13 SAMs. These trends show some similarities to those observed by \cite{cora+18} when analysing the $f_{\rm q}$ at $z=0$ for semi-analytical galaxies in the SAG model \citep{cora+06} applied on the MULTIDARK simulation MDPL2 \citep{klypin+16}. Their results for satellite galaxies are similar to ours for stellar masses larger than $10^{10} \, \msun$, while for the low mass-end, our results are between their $f_{\rm q}$ for central ($\sim 0$) and satellite ($\sim 0.4$) galaxies.
Finally, for G11, almost all galaxies, regardless of their stellar masses, are expected to have more than a 50\% chance of being quenched at $z=0$. The trend observed for G11 in the low mass-end overpredicts the results obtained for satellite galaxies in \cite{cora+18}. 

\subsection{Time evolution of galaxy properties as a function of the CG assembly channel}
We now analyse the evolution with time of the galaxy properties presented previously and their possible variation as a function of the assembly channel of the CGs. 

Given the variation described in the previous section of galaxy properties as a function of the total stellar mass of the galaxy, we split the samples of galaxy members into four subsamples according to their stellar masses at redshift zero. 

\begin{figure}
    \begin{center}
    \includegraphics[width=\hsize,viewport=0 40 440 550]{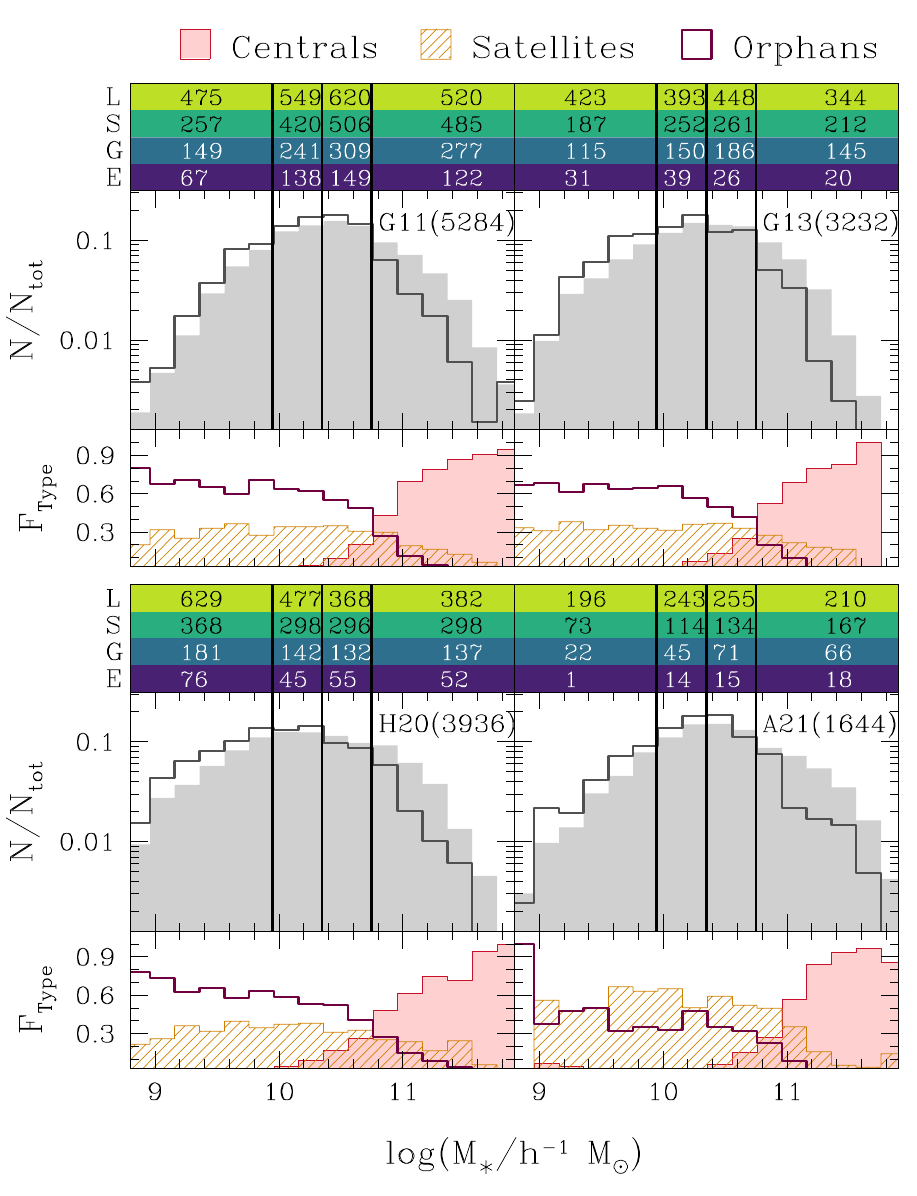}
    \caption{\label{fig:stm}
    Upper panels: Normalised stellar mass distributions at $z=0$ for CG4 galaxy members for different SAMs (\emph{shaded histograms}). Each distribution is divided into four bins using the logarithmic mass limits 9.95, 10.35 and 10.75 (vertical lines). On top of each bin of stellar mass, we quote the number of galaxies split according to their CG assembly channel (from top to bottom: \texttt{Late}, \texttt{Second}, \texttt{Gradual} and \texttt{Early)}. \emph{Empty histograms} show the normalised stellar mass distributions at $z=0$ only for the key galaxies in the CG4s.
    Lower panels: Fraction of galaxies of a given Type per bin of stellar mass.
    }
    \end{center}
\end{figure}

In the upper panels of Fig.~\ref{fig:stm} we show the stellar mass distributions at $z=0$ for galaxies in CG4s for each SAM (shaded histograms). 
We intend to define four bins in stellar mass with similar numbers of galaxies. 
Nevertheless, from this figure, it can be seen that the stellar mass distributions for G11 and A21 are slightly skewed to high masses while H20 is skewed in the opposite direction, i.e, the quartiles of the distributions are shifted from one SAM to another.
Hence, we performed a compromise decision and set the following limits in the logarithm of the stellar mass to split the four subsamples of galaxies: $\log(h\,M_*/{\rm M}_\odot)=9.95$, 10.35 and 10.75. 
These limits yield subsamples that comprise between 18\% to 32\% of the total, depending on the SAM.
Moreover, in each bin of stellar mass, galaxies are targeted according to the assembly channel of the CG they inhabit. 
The number of galaxies in each bin of stellar mass, and within a CG with a given assembly channel is quoted above the histograms of {the upper panels of} Fig.~\ref{fig:stm} (from top to bottom: \texttt{Late}, \texttt{Second}, \texttt{Gradual},  and \texttt{Early}).
However, in A21, the lowest stellar mass bin has only one galaxy from a CG with early assembly. Therefore, in the subsequent analysis, we will not show this mass bin for the early assembled CGs in the A21 SAM. 

The upper panels of Fig.~\ref{fig:stm} also show the normalised stellar mass distribution at $z=0$ of the key galaxies for each CG4 (empty histograms), i.e., the galaxy that arrived last to the quartets. 
While the key galaxies tend to be less massive than the other CG members, they contribute to as much as 12 per cent of the galaxies in the high-mass bin, compared to 25 per cent on average and 34 per cent in the lowest-mass bin.

Finally, the lower panels of Fig.~\ref{fig:stm} display the fractions of galaxies for a given type in the SAM as a function of galaxy stellar mass. One sees that, regardless of the SAM, central galaxies are the dominant type in the high-mass bin, while the low-mass bin is mainly dominated by orphans (except for A21, where we observe equal parts of satellite and orphan galaxies). The second low-mass bin is still dominated by orphans and satellite galaxies, while the third-mass bin has a mixture of the three types, with a small contribution of centrals.

\begin{figure*}    
    \begin{center}
    \includegraphics[width=0.9\hsize,viewport=20 10 480 550]{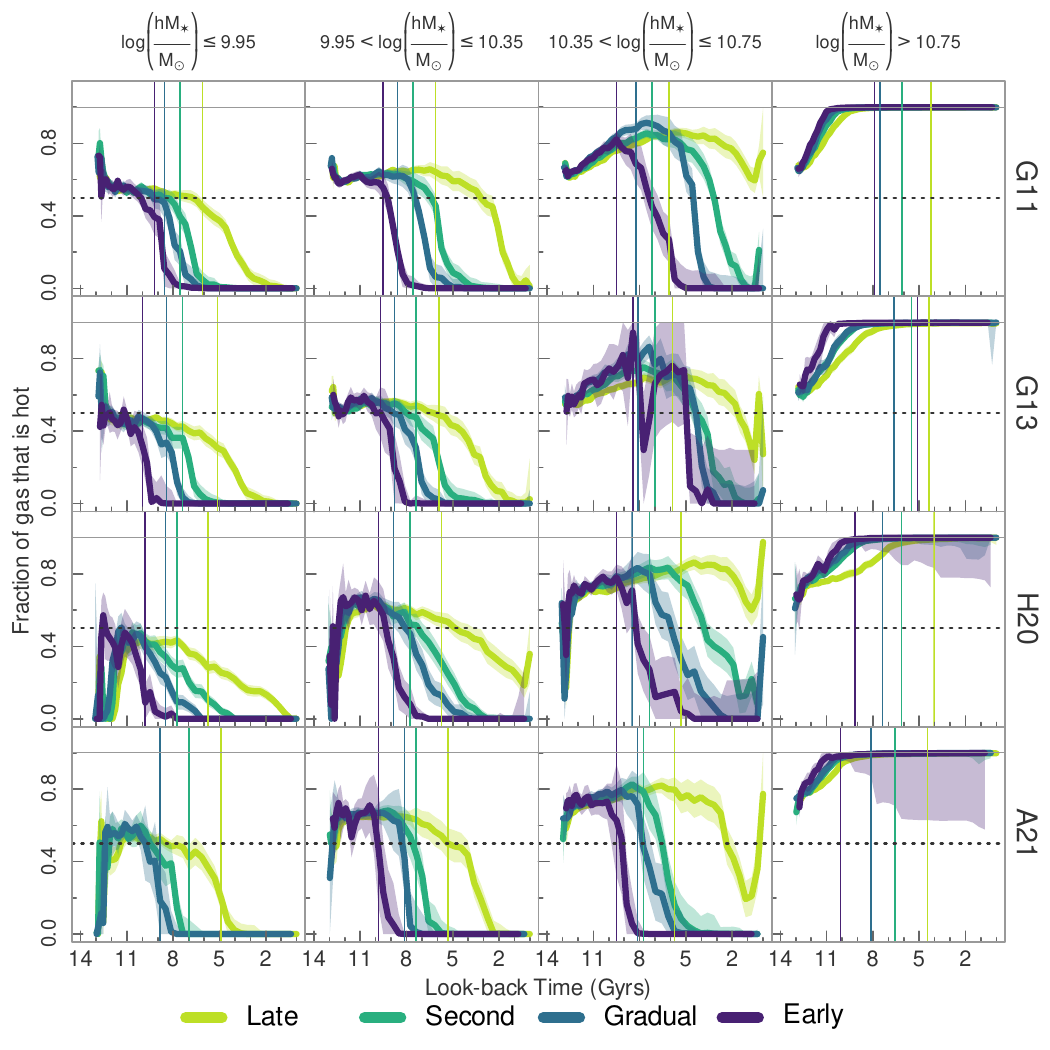}
    \caption{\label{fig:fhot}
    Fraction of gas that is hot (Eq.~[\ref{fhot}]) as a function of look-back time. The panels from left to right display the different stellar mass bins, while the panels from top to bottom display the corresponding SAM. The different curves in each panel represent the four assembly channels. Each curve value is the median of the fractions at a given time, while the \emph{shaded regions} cover the 95\% confidence interval of the median. As a reference, the \emph{dashed horizontal line} indicates a fraction of 0.5, while the \emph{thin horizontal line} marks a fraction of unity. The \emph{thin vertical lines} indicates the median's look-back times for each sample when galaxies go from being centrals to satellites (see Table~\ref{tab:times_satmed}).
}
    \end{center}
\end{figure*}

\begin{figure*}
    \begin{center}
    \includegraphics[width=0.9\hsize,viewport=20 10 480 550]{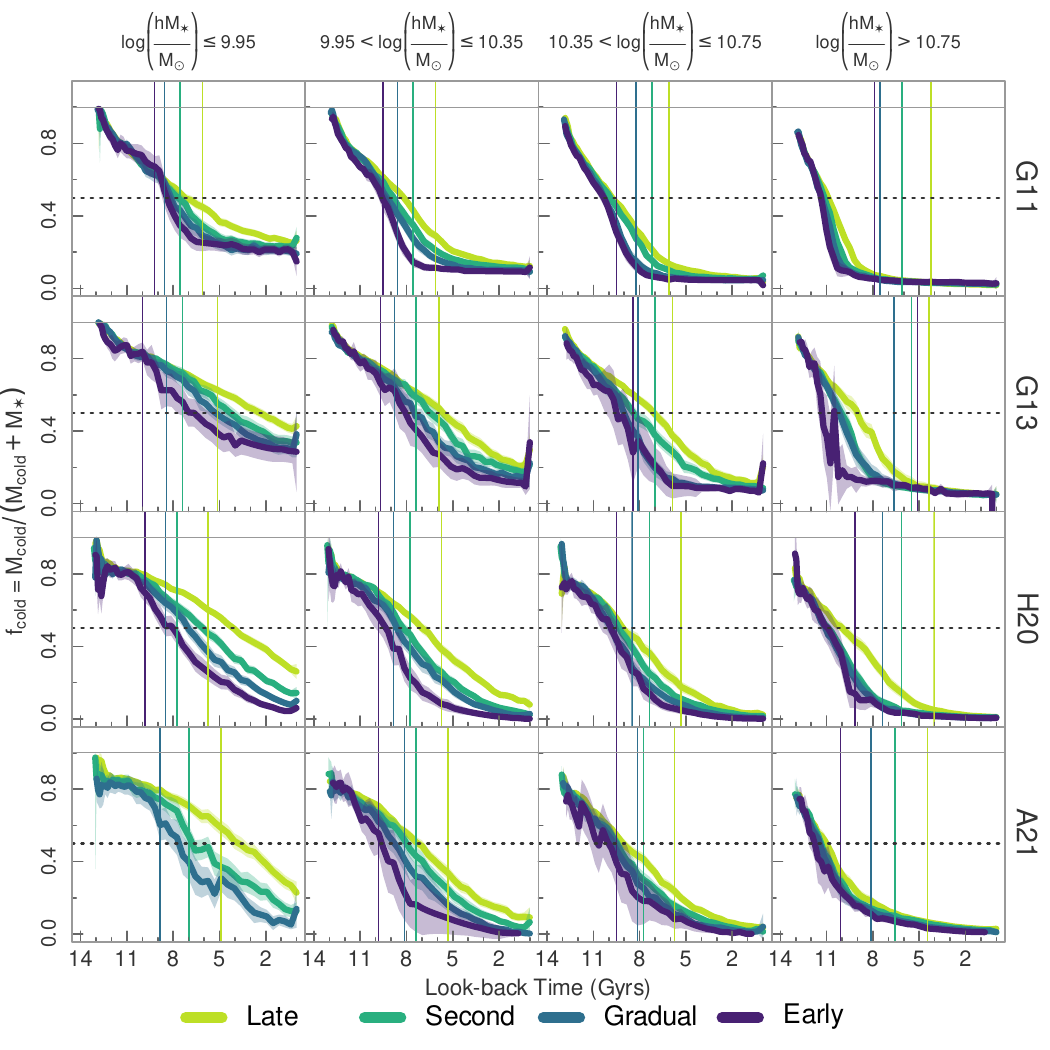}
    \caption{\label{fig:fcold}
    Same as Fig.~\ref{fig:fhot}, but for the fraction of `cold' baryons (cold gas and stars) locked up in cold gas (Eq.~[\ref{fcold}]).
     }
    \end{center}
\end{figure*}

\subsubsection{The fraction of hot gas}
We first explore the fraction of hot gas in galaxies in CGs as a function of look-back time. The adopted fraction of gas that is hot is computed for each snapshot as 
\begin{equation}
   f_{\rm hot}=\frac{M_{\rm hot}}{M_{\rm hot}+M_{\rm cold}} \ ,
   \label{fhot}
\end{equation}
where $M_{\rm hot}$ is the mass in hot gas, and $M_{\rm cold}$ as defined before. 

Figure~\ref{fig:fhot} shows the variation of the median fraction of hot gas as a function of look-back time for each galaxy subsample defined according to the $z=0$ stellar mass. Different columns correspond to different stellar mass bins, rows correspond to different SAMs, and the different colours in each panel correspond with the CG4 assembly channels.
Focusing on the dependence on the stellar mass (comparison between columns), regardless of the assembly channel or the SAM, most massive galaxies retain (on average) all their baryons in the hot gas phase over the course of their history. 
This result is not unexpected since, as stated previously, the sample of the most massive galaxies is dominated by central galaxies, which possess a vast reservoir of hot gas. Even though central galaxies undergo numerous mergers throughout their history \citep{gao+04}, their amount of hot gas will hardly be affected, as their reservoir is constantly replenished by the hot gas lost by satellite galaxies orbiting them \citep{delucia+10}.

The behaviour is different for less massive galaxies. There is a clear loss of hot gas with time, and this loss is faster for the least massive galaxies. Most of these less massive subsamples reach the present time with no hot gas, except in some cases for the sample in the third bin of stellar mass, which shows an abrupt rise in its fraction of hot gas in recent times.
This loss of hot gas is expected for these less massive galaxy samples, which are dominated by satellite and/or orphan galaxies. In older SAM versions, the hot gas was fully stripped as soon as the galaxies became satellites. 
But since G11, and therefore for all the SAMs used in this work, the recipes implemented gradual stripping of the hot gas, allowing satellite galaxies to retain a fraction of their hot gas. Hence, the gradual loss of hot gas observed in Fig.~\ref{fig:fhot} could be directly related to the time when the galaxies become satellites. 

In Appendix~\ref{sec:a0}, Fig.~\ref{fig:satellite} displays the distribution of the look-back times of galaxies when they suffer the transition from central to satellite. Vertical lines in Fig.~\ref{fig:fhot} (and also in Fig.~\ref{fig:satellite}) indicate the median look-back time values for each sample when the classification change from central to satellite (see also Table~\ref{tab:times_satmed})). 
For the stellar mass bins dominated by satellite/orphan galaxies, the look-back time medians are fairly good indicators of the time when the galaxies started to lose hot gas. Some discrepancies are observed for the lowest stellar mass bin in G13, H20 and A21, as well as for galaxies in Late subsamples. 
These discrepancies are a result of wide-ranging look-back time distributions, rendering the median an unsuitable measure (see Fig.~\ref{fig:satellite}). Therefore, in general, the patterns of hot gas loss observed in low and intermediate-mass galaxies reflect the behaviour of the satellite galaxies as they evolve over time according to the models.

\begin{figure*}    
    \begin{center}
    \includegraphics[width=0.9\hsize,viewport=20 10 480 550]{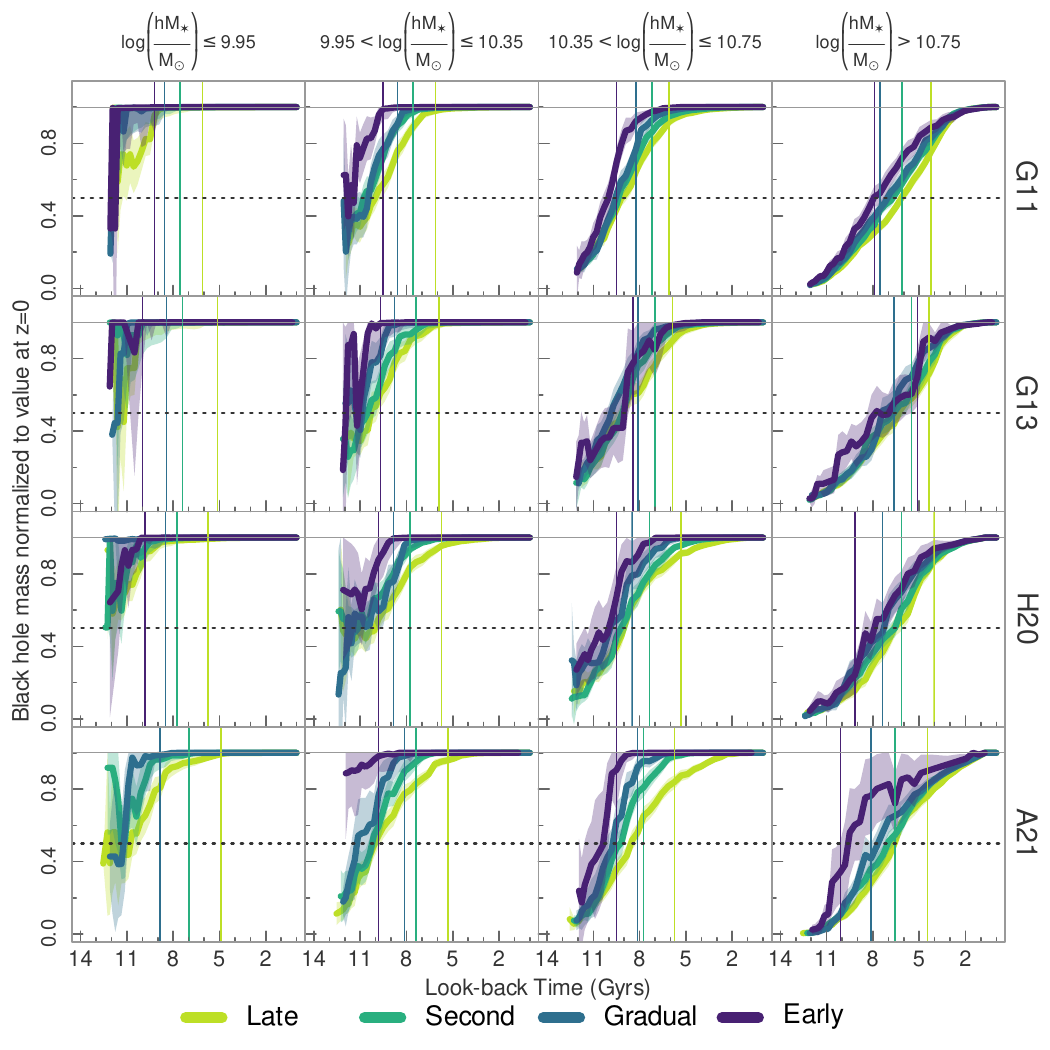}
    \caption{\label{fig:bh}
    Same as Fig.~\ref{fig:fcold}, but for the Black hole mass normalised to its value at $z=0$ (discarding the $\approx 10$ per cent of galaxies with no black hole at $z=0$, see text).
}
    \end{center}
\end{figure*}

\begin{figure*}    
    \begin{center}
    \includegraphics[width=0.9\hsize,viewport=20 10 480 550]{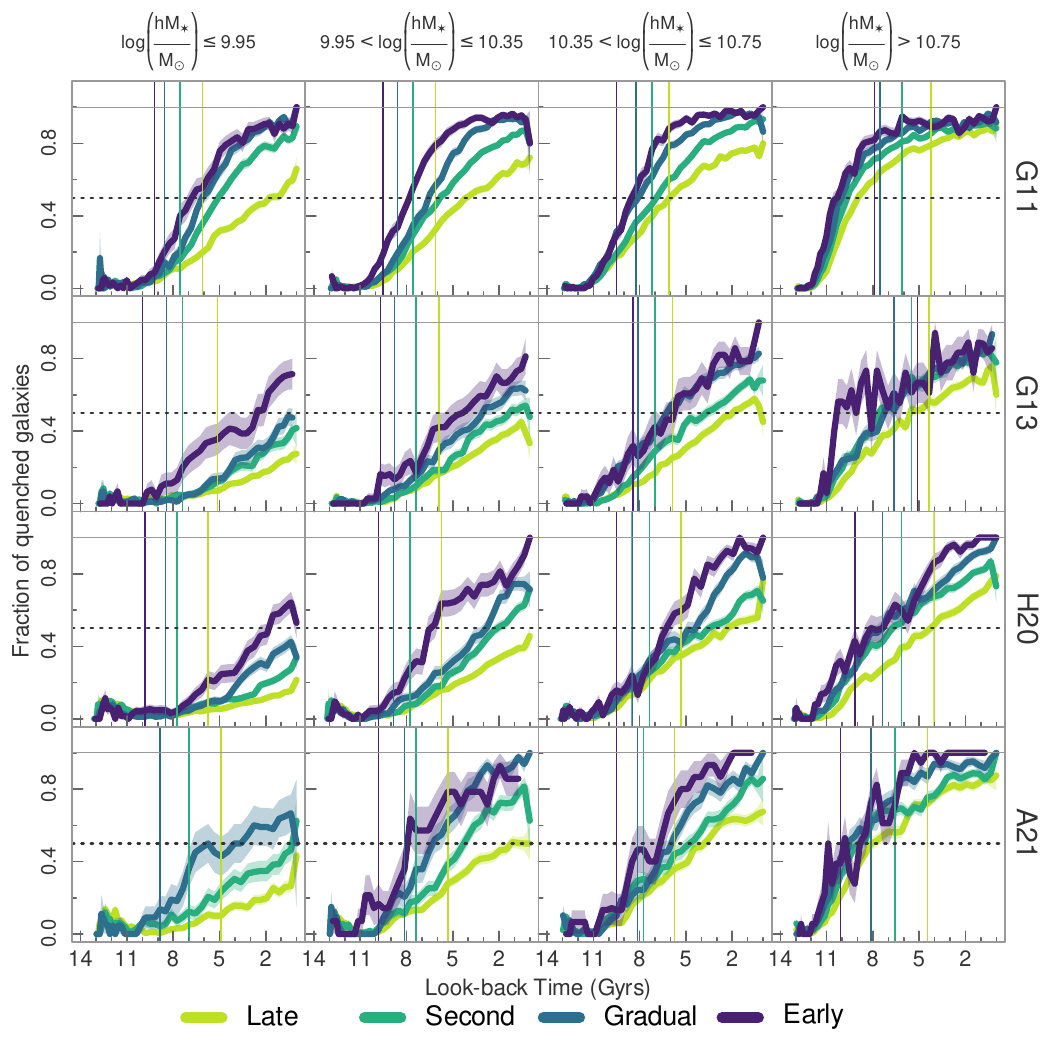}
    \caption{\label{fig:quench}
    Same as Fig.~\ref{fig:fcold}, but for the fraction of quenched galaxies. The \emph{shaded regions} cover the 68\% binomial confidence interval. }
    \end{center}
\end{figure*}

The evolution of the hot gas fraction depends quite strongly on the CG assembly channel. The hot gas fraction drops faster as we go from the newly formed CGs (\texttt{Late}) to those considered older (\texttt{Early}). In Appendix~\ref{sec:a1}, we have included a table with the look-back times when the median fractions (gas fractions, quenched fractions, and black hole mass divided by its final value) reach 0.5 ($t_{50}$, horizontal dashed lines in Fig.~\ref{fig:fhot}) for each stellar mass subsample.  Table~\ref{tab:times} indicates that the least massive galaxies that inhabit \texttt{Early} CGs lose half of their hot gas 3 Gyr earlier than their counterparts in the \texttt{Late} CGs. This behaviour is more notorious for galaxies in the third bin in stellar mass, which lose half of their hot gas almost 6 Gyr earlier in \texttt{Early} than in the \texttt{Late} CGs. 
The general dependence of the hot gas fraction can also be interpreted in terms of the satellite population. Figs.~\ref{fig:fhot}, \ref{fig:fcold}, \ref{fig:bh}, \ref{fig:quench}, and especially \ref{fig:satellite} show a correlation between the time at which galaxies shift from being central to satellite and the assembly channel. Specifically, galaxies in older systems tend to transition earlier than those that were assembled more recently.
It is worth noting that although our classification in assembly channels depends exclusively on the arrival of the last galaxy in the quartet, this does not, in principle, imply that the remaining triplet could not be a highly evolved system. However, these results could indicate that our assembly channel division has a physical implication on the evolutionary history of all the galaxies that make up these systems, and the orbit of the key galaxy is a good proxy for classification according to system formation time.

Given the low-velocity dispersions observed in CGs, the most plausible scenario indicates that tidal interactions among galaxy pairs and mergers are the most likely mechanism to act on gas, stars and dark matter \citep{boselligavazzi+06}. Clearly, tidal stripping is very important in satellite galaxies, which are the dominant type in low-mass samples. These interactions could also strip the galaxies of dark matter particles generating an orphan galaxy, the most important population in the low-mass range. This agrees with the fast declining fraction of hot gas observed over time for low-mass galaxies, since in these SAMs when a galaxy becomes an orphan lose their dark matter halo, provoking immediately the loss of its entire reservoir of hot gas. This scenario may work more efficiently in systems that have lived longer, which is clearly reflected here in the dependence of hot gas loss on the assembly channel for low- and intermediate-mass galaxies.

Finally, apart from some subtle differences, the behaviours observed for the different SAMs do not differ much. Perhaps, as a particular case, it can be mentioned that for A21 the shifting of the $t_{50}$ towards early times for earlier assembled CGs seems somewhat more extreme.

\subsubsection{The fraction of cold gas}
Secondly, we study the evolution of a proxy of the cold gas to stars ratio. 
Following \cite{cora+19} we compute this proxy as 
\begin{equation}
    f_{\rm cold}=\frac{M_{\rm cold}}{M_{\rm cold}+M_{\ast}} \ ,
    \label{fcold}
\end{equation}
where $M_\ast$ and $M_{\rm cold}$ are the running stellar and cold-gas masses.
Figure~\ref{fig:fcold} shows the variation of the medians of the fraction of cold gas as a function of look-back time for each galaxy subsample defined according to the $z=0$ stellar mass. 
 
From the comparison between columns (different stellar masses), the less massive galaxies in CG4s lose their cold gas more slowly compared to the more massive galaxies. Furthermore, the most massive galaxies have lost almost all their cold gas content by reaching $t=0$, while the least massive ones retain an average of 20\% of their cold gas. Table~\ref{tab:times} shows that more massive galaxies have loosed 50\% of their cold gas at least 3 Gyr before low-mass galaxies. 

When we subdivide systems according to their assembly channel (different colours), there is a tendency for the cold gas loss rate to be higher as the system is older. From Table~\ref{tab:times}, an increase in the $t_{50}$ times (i.e, earlier times) is observed for each mass subsample when moving from Late to Early assembly channels. This difference among the CG assembly channels is more evident in the two lowest bins of stellar mass.

In addition, we also note that different SAMs behave differently (comparison between rows). The fraction of cold gas diminishes faster for G11 galaxies. In Fig.~\ref{fig:fcold} it is observed as a very steep negative slope at early times. The fraction reaches 0.5 at earlier times regardless of the stellar mass or the assembly channel, except for the most massive bin of A21 (see Table~\ref{tab:times}). 
Also, G11 shows the smallest differences between the $t_{50}$ of the least and most massive samples ($\sim 3$ Gyr), especially for the galaxies in CG with \texttt{Late} assembly. Denser cosmologies 
show larger differences in these times for the extreme galaxy populations ($\sim 6-7$ Gyr). 

In Appendix~\ref{sec:a2}, Fig.~\ref{fig:ternary_all} provides a global view of the evolution of the 3 baryon `phases', by displaying the evolution of the fractions in stars, cold gas and hot gas for the 4 CG assembly classes in bins of final stellar mass. For the three lowest mass bins ($\log (M_*/{\rm M}_\odot)\leq 10.75$), later CG assembly leads to higher hot gas and lower cold gas without affecting the fraction of baryons locked in stars. In fact, nearly all the baryons are locked in hot gas for the third mass bin ($10.35 < \log (M_*/{\rm M}_\odot)\leq 10.75$). On the other hand, the CG assembly class has little effect on the evolution of the three fractions of baryons for the highest mass bin ($\log (M_*/{\rm M}_\odot)>10.75$), where most of the baryons are in hot gas. As previously noted, the latter result is expected, since this massive sample is dominated by central galaxies.
Fig.~\ref{fig:ternary_all} indicates that for the first three stellar mass bins in all 4 SAMs, the stellar mass fraction at a given epoch (in particular at $z=0.5$, squares) strongly depends on CG assembly history: it is much higher for early CG assembly and conversely. The trend with hot gas mass fraction is weaker and that with cold gas mass fraction is barely noticeable.

\subsubsection{The growth of the Black Hole mass}
We explore the growth of the black hole mass in each galaxy in CGs as a function of look-back time. In this analysis, the fraction of the black hole mass at a given time is normalised to its mass at $z$=0, i.e. 
$$f_{M_{\rm BH}} = \frac{M_{\rm BH}}{M_{\rm BH}^0} \ ,$$
where $M_{\rm BH}$ is the black hole mass at a given time (snapshot) and $M_{\rm BH}^0$ is the black hole mass for that galaxy at the present time ($z=0$).
However, not all galaxies have a supermassive black hole at their centre. Roughly 9, 14, 17 and 4\% of the galaxies in CG4s in G11, G13, H20 and A21, respectively, do not have a black hole, which happens more often in galaxies in the least massive bin. Therefore, for this study, we only consider those galaxies that exhibit a black hole at the present time. Figure~\ref{fig:bh} shows the evolution of the median fraction of the black hole mass as a function of look-back time. 

The evolution of the BH mass depends strongly on the stellar mass of the galaxies. The least massive galaxies in CGs tend to reach their final black hole mass at very early times, i.e., their black holes suffer few changes across their history. This behaviour changes progressively as the stellar mass of the galaxy increases, showing more gradual growth towards later times. At the opposite extreme are the most massive galaxies, which have reached their maximum black hole masses only in recent times. The $t_{50}$ gap between the least and most massive galaxies is larger than 4 Gyr (see Table~\ref{tab:times}). 

As in previous subsections, we also observed here a dependence of the fractions with the CG assembly channel, although the difference between different assembly channels is less noticeable here than for the previous fractions. In general, there is a tendency for galaxies in CGs with Early assembly to have black holes that gather 50\% of their final masses earlier than those galaxies in Gradual, Second or Late CGs. From Table~\ref{tab:times}, it can be seen that on average, galaxies in Early CGs reached their 50\% of the black hole mass $\sim 1$ Gyr earlier compared with galaxies in Late CGs, regardless of their stellar masses.   

There are small differences between SAM to SAM. The only noticeable difference occurs for the most massive subsample of galaxies in Early assembled CGs in A21 whose $t_{50}$  is larger than in the other SAMs. 

\subsubsection{The fraction of quenched galaxies}
\label{sec:quench}
Finally, we analyse the evolution of the fraction of galaxies considered passive or with suppressed SFR as a function of the look-back time. At each time, a galaxy is considered as quenched according to the definition mentioned in~\ref{sec:z0}. 

The fraction of quenched galaxies at each snapshot, $f_{\rm q}$, is shown in Fig.~\ref{fig:quench}.  
Firstly, as well known for galaxies in general \citep{wetzel+12}, the suppression of the SFR is more efficient for the most massive subsample of CG galaxies, while the least massive members of CGs display a more gradual suppression that becomes more effective near the present time. The $t_{50}$ gap between the least and most massive galaxy subsamples ranges from 4 to 8 Gyr depending on the SAM and on the assembly channel (see Table~\ref{tab:times}). 

At $z=0$, our results from the SAMs show that, in the most massive subsamples of galaxies in CGs, the fraction of quenched ones reaches 80 to 90\% regardless of the SAM or assembly channel. As suggested by previous works \citep{croton+06, martinavarro+18, Chen+20}, this particular result for massive galaxies in these SAMs is probably related to the evolution of their supermassive black holes, i.e. AGN feedback as the most likely mechanism for quenching massive galaxies. In SAMs, the black hole evolution is mainly caused by the accretion of cold gas during major mergers. Also, black hole accretion of gas from the hot atmosphere could produce powerful jets, mainly for look-back times later than 8 Gyr, causing a major impact on the final galaxy properties \citep{croton+06}. While AGN feedback can replicate the old stellar population found in massive galaxies, it is unable to match their chemical abundances \citep{delucia+12} nor the correlation between $[\alpha/{\rm Fe}]$ and the stellar mass of the galaxies \citep*{delucia+17}. Hence, the physical processes that lead to the quenching of massive galaxies in the Universe are still a topic of debate.

Except for G11, the least massive subsamples rarely have over half of their galaxies quenched at $z=0$, and therefore the $t_{50}$ is almost never reached (see Table~\ref{tab:times}). This is in agreement with their more gradual use of the reservoir of cold gas for low-mass galaxies in CGs, preserving an important amount (20 to 40 per cent) of cold gas at $z=0$. In this case, the main driving force to quench low-mass galaxies in these SAMs is probably the SN feedback. During this process, energy is released that heats up the cold gas and propels it out of the galactic disc. This can either push the gas into the hot gas component or even force it out of the subhalo and into a reservoir of ejected material. Once in this reservoir, the gas is no longer available for cooling and may never return.

Analysing the different CG assembly channels, the fraction of quenched galaxies is modulated by the assembly channel. There is a tendency for a higher fraction of quenched galaxies and a higher rate of change towards the earlier assembled CG, which becomes more evident in the less massive subsamples.

Regarding the differences between SAMs, G11 produces higher fractions of quenched galaxies and it happens earlier than all the other SAMs. 
The comparison between the G11 and G13 is very interesting since their semi-analytical recipes are almost identical. However, due to the different cosmologies, lower star formation efficiency and weaker feedback were required in G13 than in G11 to match observations. Then, galaxies in G13 are formed slightly later, shifting the peak in cosmic star formation rate to recent times, resulting in marginally more star-forming galaxies at the present.

G13, H20 and A21 look very similar to each other, with the exception of the most massive subsample for A21 which behaves more similarly to the corresponding subsample in G11. \cite{Ayromlou21} already obtained in their work a higher fraction of massive quenched galaxies at higher redshifts than the observed in \cite{Henriques+20}. The quenched fraction previously estimated by \cite{Ayromlou21} was in better agreement with observational data and is probably due to a higher AGN feedback efficiency parameter set in this SAM. Also, a more pronounced increasing rate of quenched galaxies for the low-mass galaxies in A21 can be observed when compared with H20. 
SN feedback is a significant factor in suppressing star formation in low-mass galaxies and is set differently in these two SAMs. The supernova reheating and ejection efficiency are stronger in A21 than in H20 which is probably the main cause for the different rates of quenched galaxies between A21 and H20. Another contributing factor to the increased rate of quenched galaxies may be the extension of tidal stripping to all satellite galaxies by A21. Unlike H20, this new approach is not limited to satellites within the halo virial radius. This change in the tidal stripping recipe could have played a secondary role in the measured trends. 

Finally, we conducted a test to determine if our results rely on the presence of the key galaxy in each sample. In Appendix~\ref{sec:a3}, we present Fig.~\ref{fig:quench_wkey} which displays the fraction of quenched galaxies, similar to Fig.~\ref{fig:quench}, but with the exclusion of the key galaxy in each CG. Based on the results, it seems that the key galaxy that plays a central role in the assembly channel classification has no impact on our results on CG galaxy evolution.

\section{Summary and Discussion}
\label{sec:discus}

\message{*** Discussion ***}

In this work, we present an extension of \citetalias{DiazGimenez+21} to analyse the influence of the group assembly channel in the evolution of the galaxies that populate the Hickson-like compact groups (CGs). Particularly, we selected four samples of CGs each one extracted from mock lightcone catalogues constructed from galaxy catalogues extracted from SAMs run on cosmological N-body simulations. 

Our analysis indicates that at redshift zero, low-mass CG galaxies behave differently for some properties than the full population of galaxies in the SAMs. The general trends are similar between galaxies in CGs and the full sample of galaxies, i.e. increasing function with stellar mass of the hot gas and black hole masses, black hole occupation and quenched galaxy fractions. One also sees an almost constant variation of the cold gas mass and a decreasing function with stellar mass for the specific star formation rate. 

However, as expected, the evolution of galaxies in (dense) CGs is faster than for the general population. When comparing with the entire sample of galaxies in the SAMs, there are more galaxies in CGs without hot gas for stellar masses lower than $10^{11} \,h^{-1} \,\rm {\rm M}_\odot$ (2nd row of panels of Fig.~\ref{fig:propsz0}). Additionally, those ($\lm \leq 10.5$) with a hot reservoir have a smaller amount of hot gas (top row of Fig.~\ref{fig:propsz0}). We also find that low-mass galaxies in CGs have lower specific star formation rates (5th panel of Fig.~\ref{fig:propsz0}), which leads to a higher percentage of quenched galaxies when compared to the overall sample of galaxies (bottom panel of Fig.~\ref{fig:propsz0}) in the SAMs within the same range of stellar masses.  

Our main goal in this work was to analyse whether these behaviours vary according to the assembly channel of the CG where these galaxies inhabit. 
Thus, the division of CGs according to their assembly channel shows that the behaviours described above are magnified when we go from newly formed groups to those characterised by an early assembly. In general, we observe a progressive evolution following the sequence of Late, Second, Gradual, and Early CGs. 
Galaxies in Early CGs lose their cold (Fig.~\ref{fig:fcold}) and hot gas (Fig.~\ref{fig:fhot}) content faster than their counterparts in Gradual, Second and Late CGs, with the latter showing the least extreme evolution for gas masses. Similar trends are observed for the mass growth of supermassive black holes in galaxies (Fig.~\ref{fig:bh}). Black holes seem to gain their mass faster in Early formed CGs than those located in galaxies inhabiting Late CGs. These results are in agreement with the observed trend that galaxies in Early formed CGs display a more rapid quenching of star formation than those in more recently formed CGs. 
The fraction of quenched galaxies increases faster for Early formed CGs (Fig.~\ref{fig:quench}).
These differences in galaxy evolution according to CG assembly channel are present at all CG galaxy masses, except sometimes the highest mass bin. This indicates that CG assembly affects the evolution of all their member galaxies, not just their most massive member.

These results reinforce the idea that the proposed assembly channel classification truly separates CGs into systems that have evolved along different paths. This outcome is a bit surprising since the criteria to define the different assembly channels (defined in \citetalias{DiazGimenez+21}) depend exclusively on the orbit of the latest galaxy to arrive at the compact system. At first glance, this arbitrary definition would not have to provide information on the evolution of all the galaxies in the system, especially when it is classified as recently formed since the late arrival of a galaxy does not necessarily have to define the evolution of the remaining triplet. Somehow, the time for the four CG members to converge into a truly dense group has made it possible to differentiate different evolutionary paths and, therefore, environments where the particular mechanisms that most likely intervene in the suppression of star formation can be potentiated over time.

Compact groups thus appear to be a laboratory for \emph{galaxy assembly bias}: at given stellar mass, galaxies that live in halos that formed earlier form their stars earlier. Observational results for this galaxy assembly bias were already seen in the pioneering work of \cite{Davis&Geller76}, who found that spiral galaxies (known to form stars later)  are less correlated than that of other morphological types, given that more clustered halos formed earlier \citep*{gao+05}.
On the group scale, galaxy assembly bias was previously noted in several studies:
\cite{yang+06} noticed that, at a given group mass, the galaxy-group correlation function is greater for groups with passive central galaxies, again since more clustered halos formed earlier (\citeauthor{gao+05}). 
Using the kinematics of their satellite galaxies, \cite{Wojtak&Mamon13} found that groups with red (quenched) central galaxies are (marginally) more concentrated than those of blue (non-quenched) central galaxies of the same stellar or group mass, given that, at fixed halo (group) mass, higher halo concentrations are related to earlier halo assembly \citep{Wechsler+02}.
\cite{Lim+16} found that galaxies are more quenched in groups where the dominant galaxy is more dominant, which is a sign of galaxy assembly bias since such \emph{fossil groups} are known to assemble their halos earlier \citep{Dariush+07}.
Similar signs of galaxy assembly bias within groups have been found using SAMs. \cite*{croton+07} showed that blue (later star-forming) central galaxies tend to lie in halos that assemble later.
Also using SAMs, \cite{wang+13} showed that galaxy assembly bias is present in the dependence of the galaxy clustering amplitude with the specific star formation rate, showing that low sSFR central galaxies cluster more than galaxies with the same stellar mass but higher sSFR, thus recovering the observational result of \citeauthor{Davis&Geller76}. 
Our results expand those views showing that galaxy assembly bias is also a reality for compact groups of galaxies extracted from semianalytical models of galaxy formation, reinforcing the idea of a clear link between stellar ages and group assembly time.

\section*{Acknowledgements}
{
\message{*** Acknowledgements ***}
We thank the anonymous referee for providing valuable feedback and suggestions that significantly enhanced the final version of our work. EDG and AZ have chosen to dedicate this work to the memory of their 11-year-old nephew Nahuel Brizuela Fath, a person who left an indelible mark on the hearts of those who knew and loved him, despite his brief time on this earth.

We thank the authors of the SAMs for making their models publicly available. 
The Millennium Simulation databases used in this paper and the web application providing online access to them were constructed as part of the activities of the German Astrophysical Virtual Observatory (GAVO).  
This work has been partially supported by Consejo Nacional de Investigaciones Cient\'\i ficas y T\'ecnicas de la Rep\'ublica Argentina (CONICET) and the Secretar\'\i a de Ciencia y Tecnolog\'\i a de la Universidad de C\'ordoba (SeCyT).}

\section*{Data Availability}

\message{*** Data ***}
The data underlying this article were accessed from \url{http://gavo.mpa-garching.mpg.de/Millennium/}. Galaxy lightcones built-in Paper~I and used in this work are accessed from \url{https://doi.org/10.7910/DVN/WGOPCO} \citep{data_paperI}.
The derived data generated in this research will be shared at reasonable request with the corresponding authors.

\bibliography{refs}

\appendix

\section{Transition times for satellite galaxies} 
\label{sec:a0}
\begin{figure*}
    \centering
    \includegraphics[width=\hsize]{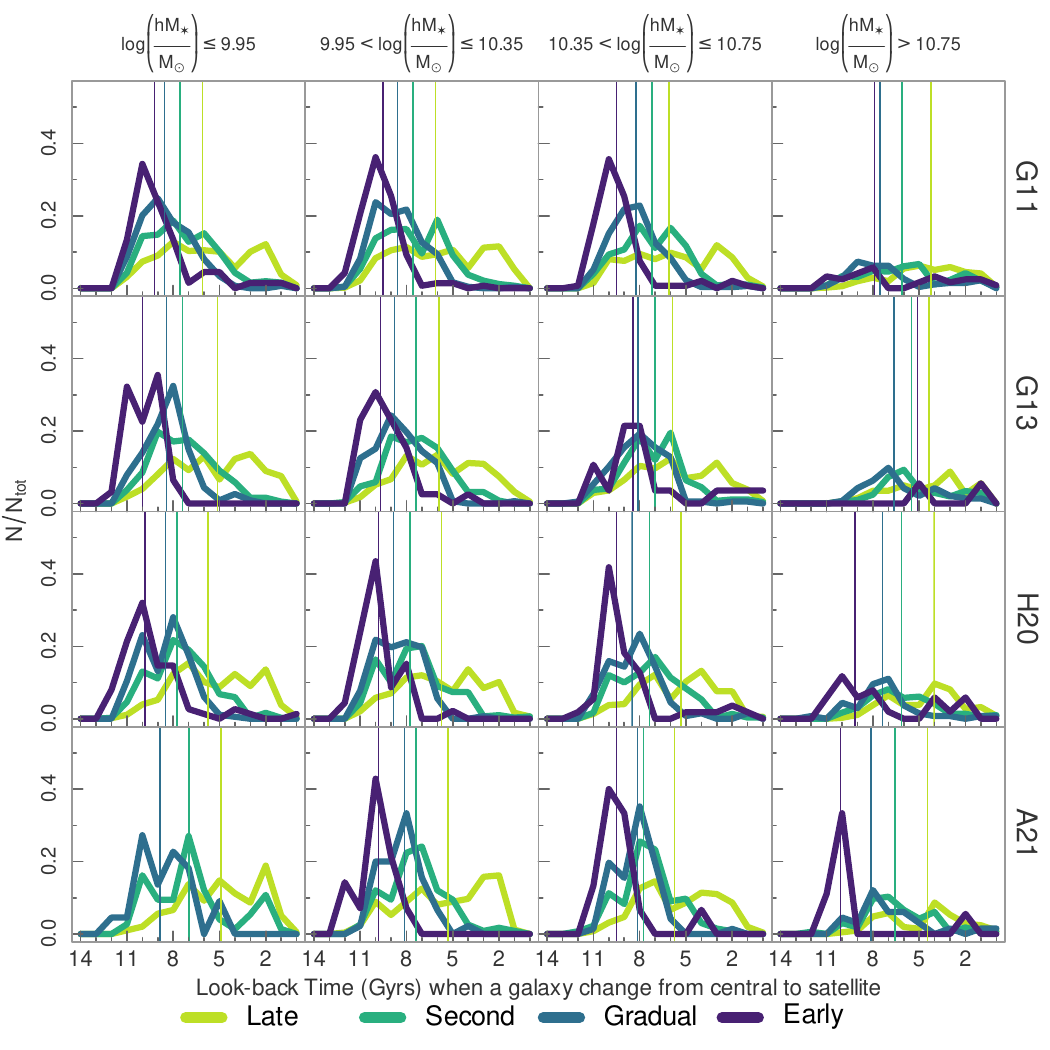}
    \caption{Distribution of look-back times of galaxies when they suffer their transformation from central to satellite.
    The panels from left to right display the different stellar mass bins, while the panels from top to bottom display the corresponding SAM. The different curves in each panel represent the four assembly channels. Vertical lines show the median values for each distribution.
}
    \label{fig:satellite}
\end{figure*}

Fig.~\ref{fig:satellite} shows the distribution of look-back times defined as the time when galaxies in CGs transition from central to satellites.

\begin{table*}
\setlength{\tabcolsep}{2pt}
\small{
\begin{center} 
\caption{Median lookback time values (in Gyr) when a central galaxy becomes a satellite (see Figure~\ref{fig:satellite}) for different SAMs, CG classes and stellar mass bins. \label{tab:times_satmed}}
\begin{tabular}{lccrrrrcrrrrcrrrrcrrrr}
\hline
\hline
 & \multicolumn{1}{c}{SAM} &\multicolumn{19}{c}{CG Assembly Channel} \\
\cline{4-22}
 & && \multicolumn{4}{c}{Late} && \multicolumn{4}{c}{Second} && \multicolumn{4}{c}{Gradual} && \multicolumn{4}{c}{Early} \\
 \cline{4-7}\cline{9-12}\cline{14-17}\cline{19-22}
  &  && $M_1$\  & $M_2$\  & $M_3$ & $M_4$\ \, && $M_1$ & $M_2$\  & $M_3$ & $M_4$\ \,  && $M_1$\  & $M_2$\  & $M_3$\  & $M_4$\  \,&& $M_1$\  & $M_2$\  & $M_3$\  & $M_4$\ \, \\
\hline
\multirow{4}{*}{Central $\to$ Satellite}
& G11 && 6.10 & 6.10 & 6.10 & 4.23 && 7.56 & 7.56 & 7.21 & 6.10 && 8.58 & 8.58 & 8.25 & 7.56 && 9.21 & 9.50 & 9.50 &  7.91 \\
& G13 && 5.12 & 5.89 & 5.89 & 4.36 && 7.38 & 7.38 & 7.02 & 5.51 && 8.44 & 8.77 & 8.09 & 6.65 && 9.98 & 9.70 & 8.44 &  5.12 \\
& H20 && 5.74 & 5.74 & 5.32 & 4.03 && 7.75 & 7.75 & 7.37 & 6.16 && 8.49 & 8.84 & 8.49 & 7.37 && 9.81 & 9.81 & 9.50 &  9.18 \\
& A21 && 4.89 & 5.32 & 5.74 & 4.46 && 6.97 & 7.37 & 7.75 & 6.57 && 8.84 & 8.13 & 8.13 & 8.13 &&   -- & 9.81 & 9.50 & 10.11 \\
\hline
\end{tabular}  
\end{center} 
\parbox{\hsize}{\noindent Notes: M1, M2, M3 and M4 respectively correspond to $\log(h\,M_\ast/{\rm M}_\odot)$ in the ranges $<9.95$, 9.95 to 10.35, 10.35 to 10.75, and $>10.75$. --: option discarded due to poor statistics.}
}
\end{table*}

\section{Characteristic Times} 
\label{sec:a1}

\message{*** Appendix ***}
Table~\ref{tab:times} displays the times when the fractions shown in Figures~\ref{fig:fcold}, \ref{fig:fhot}, \ref{fig:bh},  and \ref{fig:quench} reach 0.5.

\begin{table*}
\setlength{\tabcolsep}{2pt}
\small{
\begin{center} 
\caption{Lookback time ($t_{50}$ in Gyr) when given property reaches a fraction of 0.5 for different SAMs, CG classes and stellar mass bins. \label{tab:times}}
\begin{tabular}{lccrrrrcrrrrcrrrrcrrrr}
\hline
\hline
 & \multicolumn{1}{c}{SAM} &\multicolumn{19}{c}{CG Assembly Channel} \\
\cline{4-22}
 & && \multicolumn{4}{c}{Late} && \multicolumn{4}{c}{Second} && \multicolumn{4}{c}{Gradual} && \multicolumn{4}{c}{Early} \\
 \cline{4-7}\cline{9-12}\cline{14-17}\cline{19-22}
  &  && $M_1$\  & $M_2$\  & $M_3$ & $M_4$\ \, && $M_1$ & $M_2$\  & $M_3$ & $M_4$\ \,  && $M_1$\  & $M_2$\  & $M_3$\  & $M_4$\  \,&& $M_1$\  & $M_2$\  & $M_3$\  & $M_4$\ \, \\
\hline
\multirow{4}{*}{Cold Gas mass fraction}
& G11 &&  7.25 &  8.17 &  9.79 & 10.90 && 7.69 &  8.83 &  9.90 & 11.15 && 8.14 &  9.27 & 10.31 & 11.33 && 8.41 &  9.57 & 10.33 & 11.37 \\
& G13 &&  2.37 &  5.63 &  7.08 &  9.01 && 4.29 &  6.45 &  8.40 & 10.05 && 5.25 &  7.49 &  9.02 & 10.18 && 6.64 &  8.06 &  9.50 & 11.29 \\
& H20 &&  4.10 &  6.99 &  9.08 &  9.99 && 6.08 &  8.21 &  9.33 & 10.88 && 7.03 &  8.62 &  9.81 & 11.07 && 8.02 &  9.33 &  9.87 & 10.93 \\
& A21 &&  3.77 &  7.25 &  9.07 & 11.02 && 6.77 &  7.99 &  9.46 & 11.36 && 7.63 &  8.77 &  9.70 & 11.45 &&   -- &  9.60 & 10.02 & 11.78 \\
\hline
\multirow{4}{*}{Hot Gas mass fraction}
& G11 && 6.76 & 3.18 &  FNR & FNR && 8.63 & 6.40 & 3.18 & FNR && 9.28 & 7.45 & 4.53 & FNR && 10.10 & 9.23 & 7.38 & FNR \\
& G13 &&  FNR & 6.25 & 1.63 & FNR &&  FNR & 8.00 & 4.36 & FNR &&  FNR & 8.84 & 4.35 & FNR &&   FNR & 9.67 & 4.98 & FNR \\
& H20 &&  FNR & 4.84 &  FNR & FNR &&  FNR & 7.25 & 3.82 & FNR &&  FNR & 8.06 & 5.95 & FNR &&   FNR & 9.34 & 8.24 & FNR \\
& A21 && 8.45 & 5.32 & 2.38 & FNR && 9.50 & 7.60 & 6.54 & FNR && 9.81 & 8.28 & 7.79 & FNR &&    -- & 9.86 & 9.15 & FNR \\
\hline
\multirow{4}{*}{Black Hole mass (units of $z$=0 value)}
& G11 &&    FNR & 10.02 &  9.15 &  6.15 && FNR & 10.76 &  9.37 &  6.71 &&  11.97 & 10.46 &  9.56 &  7.05 &&  11.72 &  11.40 & 10.03 & 7.94 \\
& G13 &&  11.17 & 10.31 &  9.05 &  6.32 && FNR & 10.51 &  9.13 &  6.65 &&  11.58 & 10.87 &  9.85 &  7.42 &&    FNR &  11.93 &  9.00 & 6.97 \\
& H20 &&    FNR & 10.03 &  9.03 &  6.62 && FNR & 11.13 &  9.22 &  6.58 &&    FNR & 11.79 &  9.91 &  7.30 &&    FNR &    FNR & 10.00 & 7.91 \\
& A21 &&  11.00 & 10.04 &  8.56 &  6.60 && FNR & 10.23 &  9.24 &  7.03 &&  11.22 & 11.20 &  9.74 &  7.67 &&     -- &    FNR & 10.37 & 9.61 \\
\hline
\multirow{4}{*}{Fraction of quenched galaxies}
& G11 && 1.92 &  4.24 &  6.07 &  8.97 && 5.07 &  5.82 &  6.92 &  9.67 && 6.12 &  6.56 &  8.09 & 10.03 && 6.79 &  7.81 &  8.54 & 10.32 \\
& G13 &&  FNR &   FNR &  2.04 &  5.59 &&  FNR &  1.35 &  3.74 &  7.66 && 0.33 &  3.08 &  6.29 &  7.27 && 2.57 &  4.36 &  5.78 & 10.35 \\
& H20 &&  FNR &  0.00 &  2.32 &  4.28 &&  FNR &  2.20 &  3.56 &  6.73 &&  FNR &  2.84 &  4.38 &  7.37 && 1.84 &  6.43 &  6.27 &  7.37 \\
& A21 &&  FNR &  1.26 &  3.81 &  8.07 && 0.31 &  4.28 &  4.76 &  8.89 && 4.03 &  6.28 &  5.97 &  9.55 &&   -- &  7.94 &  6.77 &  8.84 \\
\hline
\end{tabular}  
\end{center} 
\parbox{\hsize}{\noindent Notes: M1, M2, M3 and M4 respectively correspond to $\log(h\,M_\ast/{\rm M}_\odot)$ in the ranges $<9.95$, 9.95 to 10.35, 10.35 to 10.75, and $>10.75$. --: option discarded due to poor statistics. FNR: 0.5 Fraction Never Reached. To estimate $t_{50}$ we considered the full range of look-back times, except for the Hot Gas mass where $t_{50}$ is measured between 1 and 11 Gyr.
}
}
\end{table*}

\section{Ternary plots} 
\label{sec:a2}
\begin{figure*}
    \centering
    \includegraphics[width=\hsize]{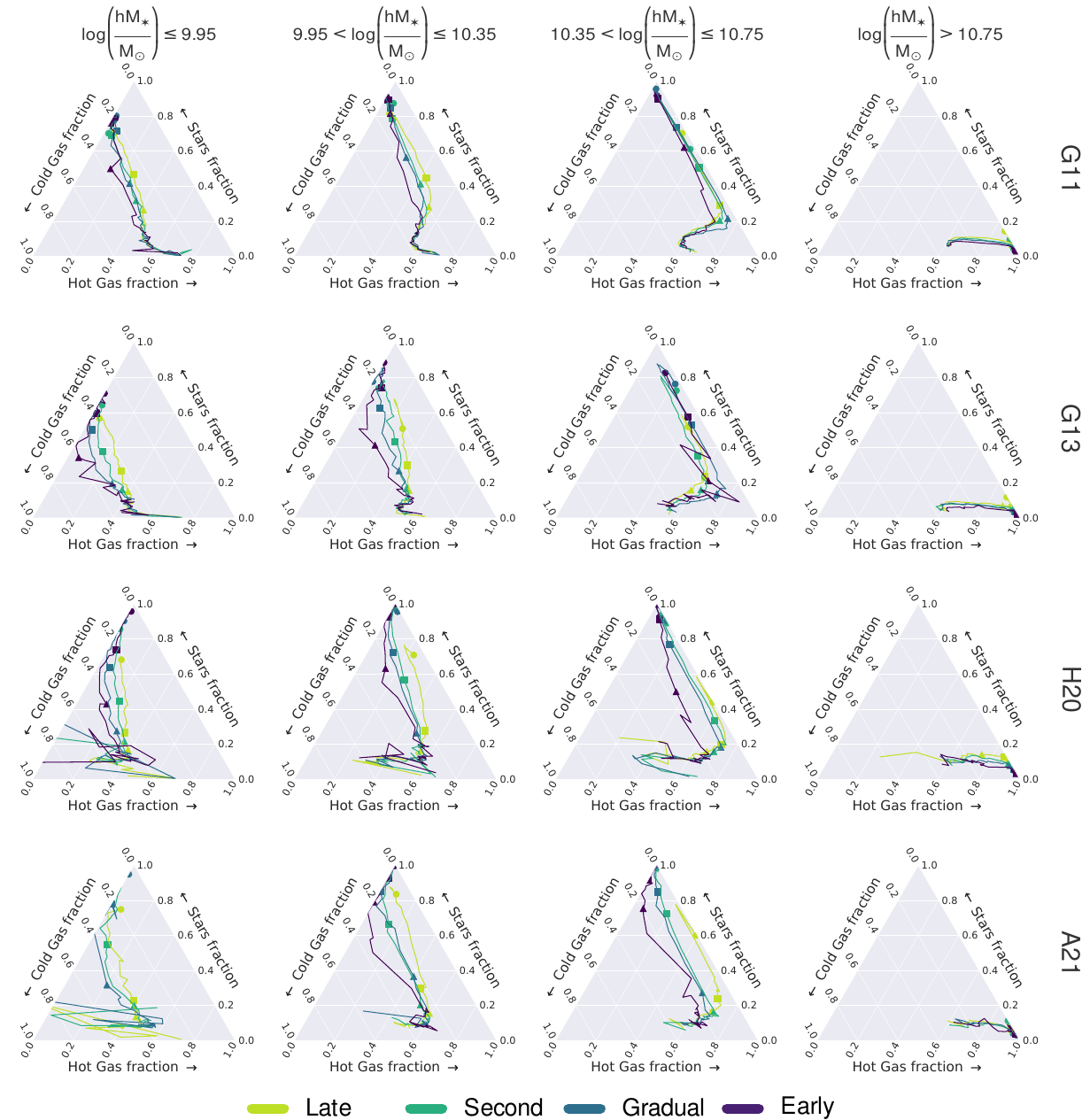}
    \caption{Ternary plots displaying all baryonic content of galaxies in CGs. The \emph{top}, \emph{left}, and \emph{right summits} correspond to all the baryons locked in stars, cold gas, and hot gas, respectively. Cosmic time follows the star fraction, i.e. increases upwards, except for the highest mass bin, where cosmic time increases towards the right, with \emph{circles} for $z=0$, \emph{squares} for $z=0.5$, and \emph{triangles} for $z=1$.  
}
    \label{fig:ternary_all}
\end{figure*}

Fig.~\ref{fig:ternary_all} shows the joint evolution of the fractions of baryons contained in stars, cold gas and hot gas per bin of stellar mass (columns) and for the different SAMs (rows). Different colours correspond to different CG assembly channels.  

\section{Fraction of quenched galaxies without the contribution of the key galaxy} 
\label{sec:a3}
\begin{figure*}
    \centering
    \includegraphics[width=\hsize]{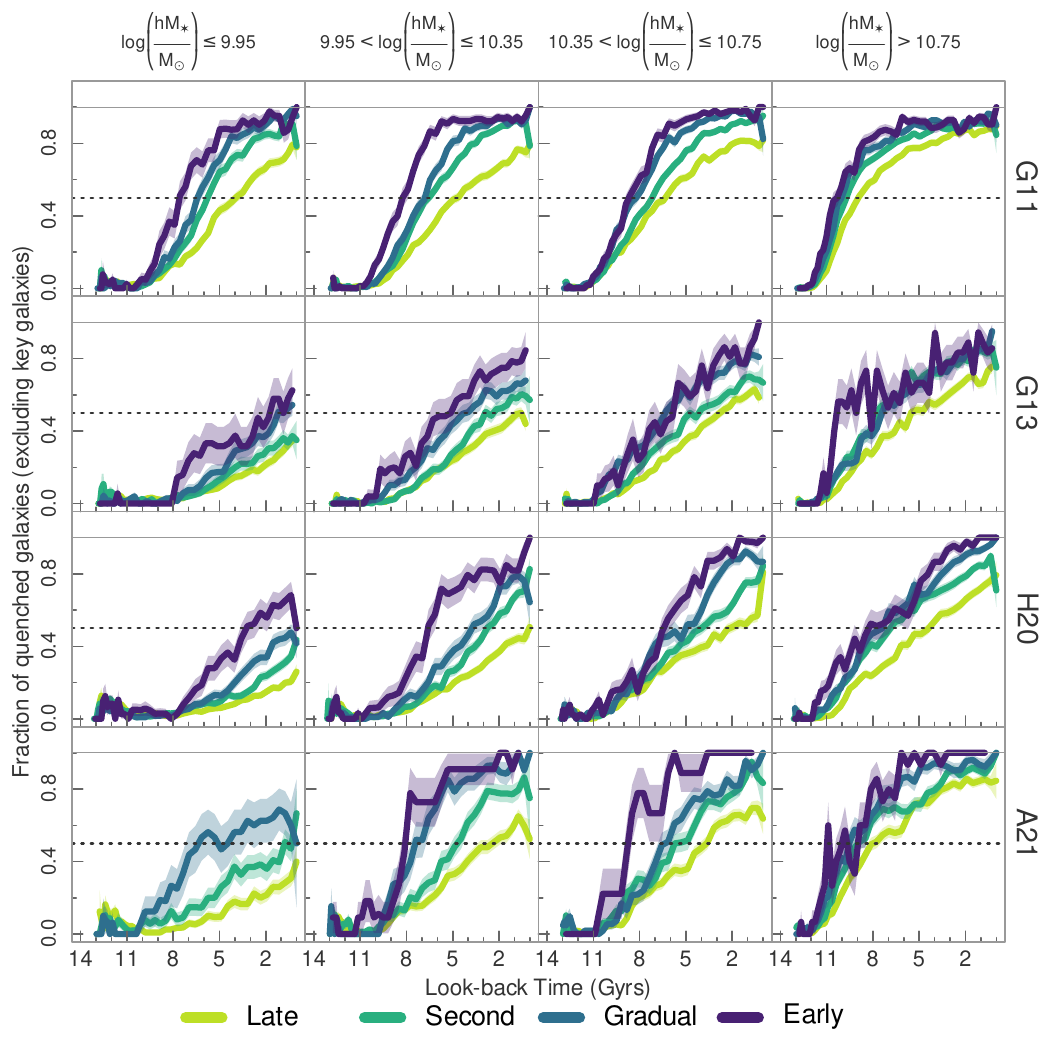}
    \caption{Same as Fig.~\ref{fig:quench}, but excluding the key galaxies from the sample. The \emph{shaded regions} cover the 68\% binomial confidence interval. 
}
    \label{fig:quench_wkey}
\end{figure*}

Fig.~\ref{fig:quench_wkey} shows the evolution of the fractions of quenched galaxies per bin of stellar mass (columns) and for the different SAMs (rows). Different colours correspond to different CG assembly channels. In this case, the fractions are calculated from samples where the key galaxies are excluded.

\label{lastpage} 

\end{document}